\documentclass[manuscript]{aastex}

\slugcomment{Accepted by the Astronomical Journal}

\shorttitle{Intergalactic Globular Clusters in the Local Group}
\shortauthors{di Tullio Zinn \& Zinn}

\begin{document}

\title{A SEARCH FOR INTERGALACTIC GLOBULAR CLUSTERS IN THE LOCAL GROUP}

\author{Graziella di Tullio Zinn and Robert Zinn}
\affil{Department of Astronomy, Yale University, P.O. Box 208101, New Haven, CT }

\begin{abstract}
The whole Sloan Digital Sky Survey (SDSS, $14,555\: \mathrm{deg}^2$)
has been searched for intergalactic globular clusters (IGCs) in the
Local Group (LG).  Using optical, infrared, and ultraviolet
photometric selection criteria and photometric redshifts, the
$2.1x10^8$ of objects in the SDSS Galaxy Catalogue were reduced to
only 183,791 brighter than $r_0 = 19$ that might be GCs.  Visual
examination of their SDSS images recovered 84\% of the confirmed GCs
in M31 and M33 and yielded 17 new GC candidates, 5 of them of high
confidence, which we could confirm as GCs in MegaPrime images from the
Canada, France, Hawaii Telescope.  These 5 GCs are within M31's halo,
but the other 12 candidates are not close to LG galaxies or galaxies
within 3 Mpc of the LG.  Even though this search covers only one-third
of the sky and some GCs could have been missed, it suggests that the
LG does not contain a large population of IGCs more luminous than $M_V
\sim -6$.  In the direction of the M81 Group, the search yielded five
candidate GCs, probable members of that group.
\end{abstract}


\section{Introduction}

This paper reports a search for globular clusters (GCs) over the whole
footprint of the Sloan Digital Sky Survey
(SDSS),$14,555\:\mathrm{deg}^2$ of the sky, out to the edge of the
Local Group (LG) and beyond in the direction of the M81 Group of
galaxies .  It encompasses a large volume of the LG, beyond the virial
radii of the Milky Way (MW) and M31 ($\sim 300$ kpc,
\citealt{garrison-kimmel14}).  There are two reasons to suspect that
GCs may be found far from the MW or M31 or any other LG group galaxy.
These intergalactic globular clusters (IGCs) may have originated in
galaxies, and through galaxy-galaxy interaction have become unbound
from their hosts and put on large orbits in the gravitational
potential of the LG \citep{bekki06}.  IGCs may have also formed
independently of galaxies in their own dark matter (DM) halos
\citep{peebles84}.  While this second hypothesis has no compelling
observational support at present, it has not been ruled out as a
source of GCs in addition to the better supported hypotheses that GCs
formed during the early evolution of galaxies (e.g.,
\citealt{harris94,kravtsov05,elmegreen12}) or from gas that was
compressed when galaxies collided (e.g.,
\citealt{schweizer87,ashman92,whitmore95}).

IGCs have been discovered in several nearby galaxy clusters such as
Fornax \citep{gregg09}, Virgo \citep{lee10}, Abell 1185
\citep{west11}, Coma \citep{peng11}, Abell 1689 \citep{alamo-mart13},
and in some of them the IGCs number in the thousands.  The majority
of the IGCs of these clusters have blue colors that are similar to the
colors of many of the GCs found in dwarf galaxies and in the blue
sequence of the typically bimodal color distribution of the GCs in
massive early-type galaxies.  The colors of these clusters are
consistent with very old ages and low metal abundances.  In the Coma
cluster, however, a significant fraction ($\sim 20\%$) of the IGCs
have been found to have red colors, suggesting that they are more
metal-rich than the blue clusters \citep{peng11}.  The common
interpretation of the presence of IGCs in galaxy clusters is that they
formed in galaxies that were later either tidally disrupted or
stripped of their outermost stars and GCs by galaxy-galaxy
interaction.  The galaxy merger Arp 105 in the cluster Abell 1185,
which is liberating both stars and GCs from their galaxy of origin
\citep{west14}, is an interaction that is producing IGCs at the
present time.  The ``hypervelocity cluster '', which appears to have
been ejected from the Virgo Cluster \citep{caldwell14}, may be
further evidence of IGC producton by galaxy-galaxy interaction
\citep{samsing15}.  Similar interactions may
have also occurred in galaxy groups according to the models of
\citet{bekki06}, and the cluster GC-2 in the M81 group, which lies
$\sim 400$ kpc from M81, may be an example of an IGC in a galaxy group
\citep{jang12}.

Galaxy interactions have occurred in the LG, because the halos of both
the MW and M31 show telltale signs of galaxy accretion.  The ongoing
disruption of the Sgr dwarf spheroidal (dSph) galaxy by the MW, and
the substructures found in the halos of the MW and M31 provide the
most direct evidence.  Other evidence that the Galactic halo was
formed by satellite accretion has accumulated for more than 30 years
\citep[see][for a review]{bland-hawthorn2014}.  The simulations of the
hierarchical picture of galaxy formation
\citep[e.g.,][]{abadi03,bullock05,zolotov09} show that the accretions
of sub-halos of DM onto a large DM halo (i.e., a large galaxy)
frequently lead to the capture of the sub-halo into a satellite orbit
and eventually its tidal disruption.  If the sub-halo is a dwarf
galaxy containing stars and star clusters, then they become part of
the stellar halo of the large galaxy.  A few of the sub-halos that
pass through the viral radius of the large halo are not captured but
remain on orbits that are several times the viral radius in extent.
Some of these so-called ``back-splash galaxies''
\citep[e.g.,][]{gill05} may have had their gas removed by ram-pressure
stripping and tidal stirring \citep[e.g.,][]{mayer01,kazantzidis13}
while passing close to the large halo.  This may explain the existence
of the Tucana and Cetus dSph galaxies in the LG that are currently far
beyond the virial radii of M31 and the MW, and yet appear to have been
stripped of their gas \citep{teyssier12}.  The recent simulations by
\citet{garrison-kimmel14} of galaxy groups containing two large halos
that are similar to M31 and the MW indicate that more back-splash
galaxies on large orbits are produced when two massive halos are
present rather than just one.  The case of the accretion of a binary
pair of sub-halos of unequal masses onto a large halo has been
simulated by \citet{sales07}, who found that often the lower mass
sub-halo was ejected at high velocity while the higher mass one merged
with the large halo.  It seems plausible then that during the
accretion of a GC-bearing dwarf galaxy by a large galaxy, one or more
of its GCs may become unbound from the dwarf and placed on large
orbits, reminiscent of the orbits of the back-splash galaxies.  Some
dwarf galaxies, for example the LG galaxies Fornax
\citep[e.g.,][]{cole12} and NGC 6822 (Huxor et al. 2013), contain GCs
that are far from their centers and therefore may be easily stripped
in a galaxy-galaxy interaction.  The large stellar substructures in the halo
of M31 \citep[and refs. therein]{fardal13,bate14} and its large number
of halo GCs, some of which appear associated with the substructures
\citep{veljanoski14} suggest that it has accreted GC-bearing galaxies.
The same is true of the MW, as indicated by the properties of its halo
GC system \citep[e.g.,][]{zinn1993,mackey04,keller12} and rather
directly by the accretion of GCs from the Sgr dSph galaxy
\citep[e.g.,][]{law10}.  It remains to be seen if the accretion events
involving the MW and M31, which theoretical simulations predict
were very numerous \citep[e.g.,][]{bullock05}, placed any GCs on such
large orbits that they are now IGCs in the LG.

The hypothesis that some GCs formed in their own DM halos is now three
decades old \citep{peebles84}, but it has not been confirmed or ruled
out observationally.  The discoveries of multiple stellar populations
and variations in the abundances of He and other elements in GCs, has
renewed the interest in this hypothesis as a way of explaining how
some GCs can retain gas and have more than one episode of star
formation \citep[see][and refs. therein]{conroyspergel11}.  Because
the DM halo of a GC may be stripped as it is accreted by a large
galaxy \citep[e.g.,][] {mashchenko05}, the most isolated and massive
GCs should be the best candidates to prove this hypothesis.  The
recent investigations by \citet{conroy11} and by \citet{ibata13} of
the structures of the GCs NGC 2419 and MGC1 in the remote outer halos
of the MW and M31 respectively, failed to find firm evidence for the
presence of DM.  The 14 dwarf galaxies that are not members of the MW
or the M31 satellite systems \citep[see][]{mcconnachie12} suggest the
presence of isolated DM halos in the LG.  In addition there might be
DM haloes in the LG that formed single GCs.  This idea and the
possibility that some galaxy interactions may have flung GCs into
larger orbits have motivated this search for IGCs in the LG, which
greatly expands our earlier one of $\sim 900 \: \mathrm{deg}^2$ of the
sky near M31 \citep[hereafter Paper II]{ditullio14}.

To our knowledge, there has not been a previous survey of a large
volume of the LG with the specific goal of finding IGCs.  Many of the
most remote GCs in the MW halo, including Pal 4 \citep{abell55} and
AM-1 \citep{madore79}, both at distances from the Sun ($d_\sun$) $>
100$ kpc, were discovered by visual examination of the photographic
plates taken with the Schmidt telescopes of the Palomar (POSS-I \&
-II), the European Southern (ESO), and the Siding Spring (U.K. Science
Research Council, SRC) Observatories.  It is hard to judge to what
distance these large surveys could have detected GCs because this
depends on the properties of the clusters, the plate material, and the
survey techniques.  Our examination of the images of several of the
brightest M31 GCs on film copies of the POSS-II plates and on images
of the digital sky survey (DSS), suggests that these plates do not
have the depth and resolution necessary to identify many GCs at the
distance of M31 ($d_\sun = 783\pm25$ kpc; all distances to LG galaxies
in this paper are from \citet{mcconnachie12}.  A good example is
provided by the luminous cluster M31 MGC1 ($M_V = -9.2$,
\citet{mackey10}, which was discovered on MegaCam images from the
Canada France Hawaii Telescope (CFHT) 3.6m telescope \citep{martin06}.
On the film copies of the POSS-II, MGC1 appears stellar.  On the SDSS
images, which are the main source of our survey technique, there is no
question that MGC1 is a GC.  Our survey can then identify similar
clusters within the boundaries of the LG.  Our survey is also
sensitive to GCs, whose central regions are not resolved in the SDSS
images, and are therefore catalogued as galaxies.  On the other hand,
it excludes GCs that are so resolved into stars that they are not
classified as galaxies by the SDSS, and therefore do not enter our
initial selection of objects from its Galaxy Catalogue, such as the MW
GCs Koposov 1 and 2 \citep{koposov07}.  Our survey misses also
ultra-faint galaxies and dSph galaxies.  Below we discuss the survey,
its application, limitations, and results.

\section{Search Area}

Figure 1, which is a Hammer projection of the sky with equatorial
coordinates, shows our search area (gray shading), the positions of
the Galactic plane, the Galactic center, M31, M33, the other satellite
galaxies of M31, the satellite galaxies of the MW, other LG galaxies,
and galaxies within 3 Mpc of the LG.  Most of these galaxies were
selected from the recent compilation of \citet{mcconnachie12}, which we
augmented with the more recent discoveries of the probable M31
satellites, Lac I, Cas III, and Per I \citep{martin13a,martin13b}
and the MW satellite, Crt I \citep{belokurov14}, which might be
instead a low-luminosity GC in the outer halo of the MW \citep{laevens14}.

In our first survey for IGCs in the LG, we surveyed $\sim
900\:\mathrm{deg}^2$ of the SDSS in an area around M31 up to 500 kpc
in projected distance from its center ($R_{gc}$) (see Fig. 1 in Paper
II), which included most of its satellite galaxies.  This area was
selected because of the possibility that some of the accretion events
that produced the large substructures and the rich population of GCs
in M31's halo, could have also produced IGCs.  Our current survey (see
Figure 1) encompasses larger expanses of the sky to the south and
southwest of M31, and it also expands in the opposite directions,
with a large gap near the galactic plane.  Note that the relatively
low galactic latitude of M31 ($-21 \fdg6$) places it almost on the
edge of the SDSS.  Since the candidate backsplash galaxies Tucana and
Cetus dSph galaxies lie far from both M31 and the MW, IGCs may be
similarly scattered throughout the LG.  A wide distribution is also
possible if IGCs formed in their own DM halos.  For both reasons it is
important to search as much of the LG as possible, and our current
search covers about one-third of the sky.

\section{Survey Techniques}

The steps in this survey were modeled after our previous surveys for
GCs in the remote halo of M31 \citep[][Paper I]{ditullio13} and for
IGCs in the vicinity of M31 (Paper II).  The first step in those two
surveys was to select objects from the SDSS Galaxy Catalogue, (a
catalogue of non-point sources), with $(g-i)_0$ colors of old GCs:
$0.3\leq (g-i)_0 \leq 1.5$.  The second step was to examine by eye the
SDSS cutout images of these selected objects as provided by the SDSS
website.  The vast majority of the objects that passed the $(g-i)_0$
color cut could be immediately rejected as galaxies.  Objects that
were not easily rejected were then examined more closely in the r
pass-band images, which we downloaded from the SDSS website in fits
file format.  In the case of the IGC survey of Paper II, which covered
a larger area than the survey in Paper I ($\sim900\: \mathrm{vs} \sim
250\: \mathrm{deg}^2$), too many ambiguous objects still remained
after this closer visual scrutiny.  Consequently, additional steps
were added in Paper II, which used combinations of optical,
ultraviolet, and infrared colors to distinguish GCs from most galaxies
on the basis of the shapes of their spectral energy distributions
(SEDs).  Although the survey in Paper II found several GCs, they are,
however, more likely to be additional members of M31's halo than IGCs
because their $R_{gc}$ values are $< 140$ kpc.  The most time
consuming part of these earlier searches was the part that involved
the initial visual inspection of the images (283,871 images in the
case of Paper II).  Because the present search area is more than 15
times larger than the one in Paper II, it was impractical to employ
the same initial technique of visual inspection.  Instead, we were
able to reject beforehand from the SDSS Galaxy Catalogue many
thousands of galaxies by using some of the photometric selection
techniques already experimented with in Paper II, based on their SED
differences with GCs.  The final step in the present survey was still
a visual inspection of the SDSS cutout images, but of a greatly
reduced number of objects.

Since the halos of the MW and M31 are probably in part the debris from
galaxy-galaxy interactions, the GCs in their halos may be
representative of IGCs.  The MW and M31 GCs have similar colors and
spectra \citep{schiavon12}, and the halo GCs in both M31 and the MW
appear to be older than 1 Gyr and metal-poor ([Fe/H]$\lesssim -1$).
Consequently, for our search we chose color ranges that isolated
similar objects.  It is not clear that these broad ranges will
necessarily include all IGCs that formed in their own DM halos because
Peebles' (1984) theory does not make specific age and metallicity
predictions beyond estimating that the formation process begins at a
redshift $\sim 50$.  We use the M31 GCs to serve as prototypes of IGCs,
because their optical, infrared, and ultra violet colors can be
measured with the same techniques that we will adopt for our survey.
We will also examine how well our methodology identifies the confirmed
GCs in the other LG galaxies with GC systems.

As we did in our two previous searches for GCs, the candidates were
drawn from the SDSS Galaxy catalogue, which contains non-stellar
objects according to the following criterion.  The SDSS frames
pipeline \citep[see][]{Stoughton02} measures the
point-spread-function (psf) of each CCD frame and uses it to compute a
``psf magnitude'' for each photometric band, to which is applied an
aperture correction to remove the effects of variable seeing.  The
pipeline also measures the light-profile of each object, which it fits
with both exponential and de Vaucouleur $r^{1/4}$ profiles after they are
convolved with the psf.  The total magnitude of an object in each band
is measured by the ``composite model magnitude'' (e.g., cmodelMag\_r),
which is based on a linear combination of the best fitting exponential
and de Vaucouleur profiles.  The test for an extended source is made
by summing over all bands the fluxes captured by the psf magnitudes
and separately the cmodel magnitudes, which are converted back to
magnitudes.  If the difference between these flux-summed magnitudes,
psfMag$ - $cmodelMag, is $\leq 0.145$, then the object is classified as a
star\footnote{http://www.sdss.org/dr12/algorithms/classify/\#photo\_class}.

Since all GCs will become indistinguishable from stars at some
distance, it is important to investigate how the SDSS
star-galaxy separation affects the completeness of our
surveys.  Figure 2 illustrates the effects of the star-galaxy
separation and our magnitude limits on our surveys for IGCs in
the LG and in the M81 group (see section 6).  In each diagram,
the open histogram is the luminosity distribution of the 168
GCs in M31, older than 1 Gyr, for which we could find data in
the ``PhotoObjAll'' catalog of the SDSS.  These objects are
confirmed GCs according to version 5 of the Revised Bologna
Catalog of M31 globular clusters (RBCv5, \citet {galleti04},
2012 edition) and/or \citet{huxor14}, and we have used the
data from these sources and, for a few objects, from the SDSS
to construct the histogram.  In projected distance from M31
($R_{gc}$), this sample spans 3.5 to 140 kpc, with 80\% at $R_{gc} > 10$
kpc.  The solid histogram in the top diagram, which includes
146 GCs, 87\% of the total, are the ones that are classified as
galaxies by the SDSS and are brighter than our survey limit
($r_0 = 19.0$, see below).  A sharp cutoff in $M_V$ is not produced
by this limit because some clusters of large radius and low
central surface brightness (e.g, HEC8 and HEC11) are measured
systematically too faint by the SDSS.  Because of this effect,
there is a bias in our surveys against finding similar
"extended" clusters \citep[see][]{huxor08}, which however
appear to be rare objects (only 4\% of the sample of 168 and 2\%
of the confirmed GCs in the RBCv5 are classified as extended).
The ragged cutoff of the solid histogram is also caused by our
$r_0$ limit depending only on the foreground MW extinctions of
the objects, while the extinctions used to compute $M_V$ include
contributions from M31.  In addition, 3 of the 168 clusters in
the sample fail the star-galaxy separation and were listed as
stars in the PhotoObjAll catalog. 

The solid histogram in the middle diagram, which contains 131 clusters
or 78\% of the total number, includes the effects of the $r_0$ limit
and star-galaxy separation at 1100 kpc from the Sun.  This may be
approximately the limit of the LG in most directions of our survey
because McConnachie (2012) estimated that the zero velocity surface of
the LG lies $1060\pm70$ kpc from the mid point between the MW and M31.
To model the SDSS star-galaxy separation, we used the composite
profile fit in each band to find the radius of the aperture that
enclosed the same flux as the psfMag.  This radius was typically 80\%
of the full-width-half-maximum (FWHM) of the psf as listed in
PhotoObjAll.  We then scaled the effective radii of the exponential
and de Vaucouleurs profiles that are listed in the PhotoObjAll catalog
from the M31 distance to 1100 kpc and computed new composite profiles
for each band.  Using these profiles and the effective apertures of
the psfs, we computed the difference psfMag$ - $cmodelMag, which, if
$\leq 0.145$, indicated a stellar source.  According to this
procedure, 10 of the 168 clusters would be considered as stars at a
distance of 1100 kpc, but only 2 of them are brighter than the $r_0$
cutoff.

The solid histogram in the bottom diagram of Figure 2
shows the effects of the $r_0$ cutoff, now increased to 20.0 (see
section 6), and the star-galaxy separation at the distance of
the M81 Group, 3.6 Mpc (Karachentsev et al. 2013).  Only 35
clusters or 21\% of the original 168 are included in this
histogram.  Nonetheless, it suggests that $\sim 50\%$ of the most
luminous GCs ($M_V \leq -7.8$) are non-stellar at M81's distance.
This result is in qualitative agreement with the presence of
some M81 GCs in the SDSS galaxy catalogue (e.g., the two
studied by \citet{jang12}).

For the optical data, we downloaded from the Galaxy catalog the SDSS
"model magnitudes", which are based on the better one of the
exponential or the de Vaucouleur fits to the r-band light profile.
While the model magnitudes provide the best measurements of the colors
of galaxies, they are less good than the cmodel magnitudes for
measuring the total
light\footnote{www.sdss.org/dr12/algorithms/magnitudes/}.  Over the
magnitude ranges of interest here the differences between these
magnitudes are so small that they can be safely ignored (e.g., for the
147 M31 clusters with $r_0 \leq 19.0$ in Fig. 2, the difference,
modelMag$ - $cmodelMag, have means of $0.031\pm0.010$,
$0.016\pm0.005$, and $0.008\pm0.009$ in the g, r, and i bands).  We
only selected objects with reddening corrected r magnitudes $12.0 \leq
r_0 \leq 19.0$, because our experience with the M31 GCs suggested that
fainter ones might fail our color selection criteria, and with very
few exceptions, cannot be confirmed as star clusters by visual
inspection of the SDSS images.  For an optical color we chose $ g-i $
because it is well measured for these relatively bright objects and
has a long wavelength baseline.  In addition this time we selected for
our sample objects with reddening corrected colors in a more
restricted range, $0.3 \leq (g-i)_0 \leq 1.1$, which still encompasses
the range in color of the outer halo GCs in M31 and most of the GCs
belonging to the dwarf galaxies of the LG (see section 4).  Because of
the age-metallicity degeneracy effect, this range in color encompasses
wide enough ranges in both metallicity and age,$ -2.25 \lesssim
\mathrm{[Fe/H]} \lesssim -0.33$ if age $> 1$ Gyr, according to the
models of \citet{maraston98,maraston05}.  Moreover the selection
criteria for GCs that we developed in Paper II, using color
combinations from the SDSS, the Wide-field Infrared Survey Explorer
(WISE), and the Galaxy Evolution Explorer (GALEX) satellites, work
well within this range of $(g-i)_0$, but not outside it, as shown in
Figure 3.  These criteria were presented and discussed at greater
length in Paper II.  Because the recent Data Release 10 (DR10) of the
SDSS Galaxy Catalogue is linked to the WISE All-Sky Catalogue, much of
our present survey could now be done automatically via Casjob on the
SDSS website.

Our previous experience in searching for GCs in the halo of M31 has
shown that the largest contamination in our sample comes from galaxies
resembling GCs in appearance in the SDSS images because they look
compact and nearly round. When the spectroscopically measured
redshifts (z) of these galaxies were available, they indicated that
the galaxies lie far beyond the LG.  In most cases, the SEDs of these
galaxies do not closely resemble those of GCs.  Unlike GCs, many of
them are still forming stars.  Ones that have little or no star
formation have stellar populations that are more metal-rich, and hence
redder than the GCs found in the halos of the MW and M31 or in dwarf
galaxies.  Therefore these groups of galaxies can be separated from GCs
by means of photometric criteria. To illustrate how our selection
criteria can distinguish between these galaxies and GCs, we
have selected a sample of 122 GCs in M31 with $R_{gc} \geq 10$ kpc and
$12 \leq r_0 \leq 19$, that do not lie in close proximity to bright stars or
in the very dense star fields of M31.  These confirmed clusters were
selected from the catalogues of \citet{kang12,huxor08,huxor14}, and
Papers I and II.  Because we are searching for IGCs, in the
construction of the selection criteria we selected a more remote
sample ($R_{gc} \geq 10$ kpc) than the one used in Paper II ($R_{gc}
\geq 3$ kpc).  The clusters in the inner regions are more likely to
have formed within M31 rather than in dwarf galaxies or in their own
dark matter halos.  In the diagrams of Figures 3 and 4, this
sample of GCs is compared to the objects in a small area of our survey
covering $4 \: \mathrm{deg}^2$, and centered at RA, Dec = 161.0, +1.0
(l = +249.1, b = +50.1).  This region is representative of all but the
lower galactic latitude regions of our survey, where the SDSS Galaxy
Catalogue contains fewer galaxies per square degree and many more
tight groups of stars.  In this comparison region, the SDSS Galaxy
Catalogue contains 2146 objects with $12.0 \leq r_0 \leq 19.0$.  The
SDSS cutout image of each of these objects was visually examined and
none of them was a candidate GC according to the criteria that we
discussed in Paper I.  We used these objects to illustrate how the
majority of similar objects can be separated from GCs without
resorting to a preliminary visual inspection, which would have been
very time consuming for the area of $14,555\: \mathrm{deg}^2$ surveyed
in this search.

The top diagram of Figure 3 shows that many galaxies (X's) are redder
in the $(g-i)_0$ and $(i-W1)_0$ colors than GCs (circles). The color
$(i-W1)_0$ is formed from the magnitudes in the SDSS i-band,
transformed to the Vega system, and the WISE W1 band, which is
centered at $3.4 \micron$, while the W2 and W3 bands are centered at
$4.6$ and $12 \micron$.  We used the profile fitted magnitudes for all
WISE measurements because they are presented as the most reliable
measurement for unresolved sources in the description of the WISE
catalogue \citep{cutri11}.  Nearly all of the M31 GCs are unresolved
by WISE in the W1, W2, and W3 wavelength bands, which have psfs with
full-width-half-maxima (FWHM) of $6\farcs1, 6\farcs8$, and $7\farcs4$,
respectively.  Later we will discuss the effects on our survey of
forming colors with the SDSS model magnitudes, which include nearly
all of the light of the objects, and the WISE and GALEX measurements,
which do not.  The top diagram in Figure 3 shows, in the $(g-i)_0 -
(i-W1)_0$ plane, the locations of the objects of the comparison region
(X's), which extend beyond the limits of this plot, and of the
selected sample of M31 GCs (circles).  The dashed contour in this
diagram encloses 90\% of the density of the 2-D kernel density
estimate of the GCs.  The rectangular area, which encloses 120 (98\%)
of the GCs, marks some of the cuts that we will impose to isolate GCs,
$0.3\leq (g-i)_0 \leq 1.1$ and $(i-W1)_0 \leq 2.03$.  Only 389 (18\%)
of the objects in the comparison region pass these cuts.

The middle diagram in Figure 3 shows the usefulness of adding a second
cut with the photometric redshift, Photoz, which is derived from the 5
SDSS photometric bands by the K-D tree method and listed in DR10.  In
Paper II, we showed that Photoz is a useful discriminant between M31
GCs and galaxies because it systematically assigns low values of z to
the GCs (see Fig. 4 in Paper II).  Although these Photoz values are
much larger than the real z's of the GCs, they are smaller than the
ones of many galaxies of similar color and magnitude.  For this study,
we chose for Photoz a wider limit ($\leq 0.13$) than in Paper II to
reduce a bias against fainter GCs, like B407 which would not pass
otherwise our selection. The original sample of 122 M31 GCs had to be
reduced to 89 in the middle diagram of Figure 3, because values of
Photoz were available for only the ones within the SDSS footprint.
One of these clusters, B517, has an exceptionally large value of
Photoz, 0.177, and it was excluded when calculating the 2-D kernel
density and its 90\% contour.  The $(u-g)_0$ color of B517 is
approximately 0.2 mag smaller than the ones of other M31 GCs of
similar $(g-i)_0$, which may be responsible for its anomalously large
Photoz.  The rectangle in this diagram indicates the region delineated
by the cuts in Photoz and $(g-i)_0$ that we will use to isolate GCs,
and it encloses every GC (circles) except B517.  The 389 objects in
the comparison region that passed the cuts in the upper diagram are
the ones plotted in this middle diagram; 338 of them passed the Photoz
cut.

The bottom diagram in Figure 3 is a plot of the color W2-W3, versus
$(g-i)_0$.  W2-W3 is sensitive to the amount of star formation because
the W3 band includes the emission from polycyclic aromatic hydrocarbon
molecules. While all 122 M31 GCs are plotted in this diagram, W3 was
measured for only 31 (solid circles).  For the other 91 GCs, W3 is an
upper limit (open circles), and therefore W2-W3 is also an upper
limit.  Using only the GCs with measured values, we calculated the 2-D
kernel density, and the dashed contour in the diagram encloses 90\% of
the density of these GCs.  The faintest GCs are the ones lying above
the contour.  In order not to exclude faint GCs with only upper limit
values, we will impose two different cuts by W2-W3.  If W2-W3 was a
measured value, like for solid circle objects, the cut will be at
3.1.  If W2-W3 was an upper limit, it is raised to 4.0.  The majority
of M31 GCs ($\sim96\%$) passed all of these three criteria.  Also
plotted (crosses: upper limit; X's: measured value of W2-W3) are the
338 comparison objects that passed the previous two criteria.  Only
128 of them or 6\% of the original 2146 objects passed the final cut,
demonstrating the value of this sequence of selection photometric
criteria to reduce galaxy contamination.

This selection procedure through WISE can be largely automated because
the SDSS Galaxy Catalogue is linked to the WISE All-Sky Catalogue.
Therefore, the first step for our selected survey for GC in the whole
SDSS footprint was to query the SDSS Galaxy and the WISE All-Sky
catalogues, via CasJobs on the SDSS website, for objects with $12.0
\leq r_0 \leq 19.0, 0.3 \leq (g-i)_0 \leq 1.1 , (i-W1)_0 \leq 2.03,\,
\mathrm{Photoz} \leq 0.13,\: \mathrm{and} W2-W3 \leq 3.1 $ for
measurements of W2-W3, or $ \leq 4.0 $ for upper limits.  Since the
footprint of the SDSS contains $\sim 2.1x10^8$ galaxies of all
magnitudes and colors, a huge number of objects ($\sim 4.21x10^5$)
passed our first five selection criteria, still a sample too large to
be inspected visually in a reasonable time, and still containing a
large component of galaxies.  At this point we introduced another cut,
also discussed in Paper II, based on the color $(NUV-g)_0$, which
demonstrated to be very useful for separating GCs from galaxies with
large UV fluxes, presumably because they contain many hot and luminous
young stars. NUV is the GALEX magnitude in its near UV band ($\lambda$
effective $ = 2271$ \AA) and g is the SDSS g-band magnitude.  This
additional criterion is illustrated in Figure 4 with the same sample
of M31 GCs previously used (open circles) and the objects in the
comparison region (crosses and X's) that passed the previous cuts, and
could also be identified in the Galex catalogue.  The NUV measurements
for many of the GCs were taken from the catalogue of \citet{kang12}.
For the remaining clusters and for the whole samples of comparison
galaxies, we compared their positions with the objects in the GALEX
catalogue that is available through the GALEXview website and
considered a match if their positions agreed to $\leq 6\farcs0$.  The
average difference between the SDSS and GALEX positions was only
$1\farcs4$, with $< 10\%$ having differences $>3\arcsec$.
Consequently, they are smaller than or, in the worse cases, comparable
to the NUV psf ($5\farcs3$ to $\sim 8\arcsec$, \citet{morrissey07}).
These results and the relatively low density of GALEX sources on the sky
\citep[e.g.,][]{bianchi14}, suggest that very few of the objects
drawn from the SDSS catalog were matched with the wrong GALEX source.
If more than one GALEX measurement was available for an object, we
computed the weighted average, using as weights one over the square of
the error listed for NUV.  Using the sample of 122 GCs, we computed
the 2-D kernel density and the dashed contour in Figure 4 encompasses
90\% of the density.  The curve that separates most of the comparison
objects from the GCs is identical to the one plotted in Figure 2 of
Paper II, which had a larger sample of galaxies.  While all 122 GCs
lie below this curve and therefore pass this last criterion, only 16
of the remaining comparison objects lie below, and none of them could
be classified as candidate GCs by visual inspection.  After these
tests, we concluded that the combination of the selection criteria
illustrated in Figures 3 and 4 could reduce the number of objects to
be visually examined to a small percentage of the initial sample
($\sim 1\%$), while retaining most of the GCs (96\%), making feasible
our planned survey of the whole SDSS footprint.  Therefore the second
step, based on the color $(NUV-g)_0$ of our procedure was to query the
GALEXview website for GALEX sources that matched the positions of the
objects ($\sim 4.21x10^5$) that passed the WISE and Photoz cuts
introduced in the first step of the survey. We found approximately
$3.17x10^5$ coincidences, and to these objects we applied the cut
based on the $(NUV-g)_0$ color (Fig. 4). The objects that passed this
last cut, added to the ones that we could not find in GALEX, formed
our final sample of 183,791 objects to be visually inspected.

\section{Application to the Globular Clusters in other Local Group Galaxies}

The above criteria appear to be well tuned for separating the halo GCs
of M31 from galaxies, but do they also work for the GCs in the
smaller galaxies of the LG, which may be better analogues of the
IGCs?  To examine this question, we compiled similar data on the GCs
in the Fornax dSph galaxy, a satellite of the MW, in four satellites of
M31 (M33, NGC 205, 147, 185), and in two dwarf irregular galaxies
of the LG, NGC 6822 and Wolf-Lundmark-Melotte (WLM).  Because most of
these galaxies do not lie in the SDSS footprint, we could not include
measurements of Photoz in this discussion.  Moreover we did not
consider the GCs in the MW, the Magellanic Clouds, and the Sagittarius
dSph galaxy because they subtend such large angles on the sky that
their measurement would require substantially different techniques
than we employ in our survey of the LG.

In the case of the Fornax dSph galaxy ($d_\sun = 147 \pm 12$ kpc), its
GCs subtend moderately large angles on the sky, with $R_h$ values of
$4\farcs7, 6\farcs4, 7\farcs0, 10\farcs9,$ and $18\farcs7$ for GCs F4,
F3, F5, F2, and F1, respectively \citep{mcLaughlin05}.  They provide a
good test of the effects of using magnitudes, such as the WISE and GALEX ones,
which do not capture all the light of an object.  We computed the
$(i-W1)_0$ and W2-W3 colors using both the profile-fitting WISE
magnitudes, which we adopted for our survey, and the largest WISE
aperture, which is $24\farcs75$ in diameter.  For the Fornax GCs, we
could not find in the literature measurements in the g or i
pass-bands.  Consequently, we made estimates that are based on the V
and B-V measurements of F2, F3, F4, and F5 by \citet{gordon83}, on the
values compiled for F1 by \citet{harris79} and the reddening from
\citet{mackey03}. We then constructed a transformation between
$(B-V)_0$ and $(g-i)_0$ from measurements of M31 GCs in both colors.

In the diagrams of Figure 5 are plotted the Fornax GCs and the GCs in
the other LG galaxies of our sample. The contours are the ones
calculated for Figures 3 and 4.  For each Fornax cluster, the colors
formed by the profile-fitted WISE mag (filled circles) and the largest
WISE aperture (X's) are plotted and joined by dashed lines.  Given the
large $R_h$ of F1, it is not surprising that with the profile-fitted
W1 magnitude, $(i-W1)_0$ lies far outside the 90\% contours that are
defined by the sample of M31 GCs.  F1 nonetheless passes our
$(i-W1)_0$ criterion.  Despite the fact that the $R_h$ of F1 is
significantly larger than the psfs of W2 and W3, its W2-W3 color
passes the cut when the profile fitting magnitudes are used.  The
results obtained for other Fornax GCs with profile-fitting magnitudes
are less deviant than the ones for F1, and with the largest WISE
apertures all 5 Fornax clusters lie within or very close to 90\%
contours of the M31 sample.  In the plot of $(NUV-g)_0$ vs. $(g-i)_0$,
all 5 Fornax clusters lie within the 90\% contour despite their
relatively large angular sizes.  Because Fornax is well within the
boundary of the MW's halo as defined by its satellite galaxies (see
McConnachie 2012), F1 may be larger in angular size than any IGC in
the LG, unless they are atypical GCs.  None of the GCs in the more
distant LG galaxies have $R_h > 10\arcsec$ (see refs. below).
The test provided by the Fornax clusters in Figure 5 suggests that the
mismatches in the amounts of light included in the SDSS model
magnitudes and the WISE and GALEX magnitudes are unlikely to cause GCs
to be missed by our selection criteria.

Recent surveys of the dwarf irregular galaxy NGC 6822 ($d_\sun =
459\pm 17$ kpc) by \citet{hwang11} and \citet{huxor13} have
added 7 new GCs, bringing the total known to 9.  We could reliably
place 6 of these clusters (HVII, C2, C3, C4, SC6, SC7) in the
diagrams that use WISE measurements and 4 of them (C2, C3, C4, SC7)
in the one that uses GALEX (solid triangles in Fig. 5).  The other
clusters are too contaminated by a rich star field (HVIII), a nearby
bright star (C1), or simply too faint (SC5) to be measured by WISE and
GALEX.  The identification of HVII and SC6 in the GALEX catalog was
problematic because of other nearby sources.  The above references
provided the g and i-band photometry and the reddening values that we
used.  For HVII, we estimated $(g-i)_0$ from the $(V-I)_0$ color that
was listed by Hwang et al. (2011).  The reddening is variable in the
direction of NGC 6822.  \citet{hwang11} quote values of $E(g-i)$
from 0.16 to 0.36 for the clusters in their study.  The $(g-i)_0$
color of C3 \citep{hwang11} is exceptionally red (1.35) for its
metal-poor composition([Fe/H] = -1.61), which \citet{hwang14}
measured from its spectrum.  C3 failed every one of our GC criteria
because of its red color.  Our survey technique would have discovered
5 of the 9 clusters in NGC 6822, (or 56\%).  This limited success
should not be indicative of our search for IGCs in the LG because some
of the failures and omissions of the NGC 6822 clusters were caused by
contamination from other sources in this galaxy, and presumably
IGCs are in less dense fields.

The other LG dwarf irregular galaxy with GCs is WLM ($d_\sun =
933\pm34 $ kpc), which contains one luminous GC ($M_V \sim -8.9$).  We
estimated its $(g-i)_0$ from the $(V-I)_0$ color measured by
\citet{billett02}.  It is identified unambiguously in the WISE
catalog, and passed our criteria based on WISE measurements (open
circle in Fig. 5). It was not found by GALEX, presumably because of
its proximity to other UV sources.

M33 ($d_\sun =809\pm22 $ kpc), the spiral galaxy companion of M31,
lies within the SDSS footprint, although few of its GCs are in the
SDSS Galaxy catalog because they are projected on the dense star
fields of this face-on galaxy.  Using the lists of
\citet{sarajedini07,huxor09,cockcroft11}, we identified 10 GCs in the
SDSS Galaxy catalog with reliable photometry, 7 of which are brighter
than the $r_0$ cutoff of our survey and passed also our Photoz
criterion.  All 7(crosses in Fig. 5) passed the $(i-W1)_0$, and all
but one (S in \citealt{cockcroft11}), passed the W2-W3 one. Five of the
seven clusters were identified in the GALEX catalogue, and all of them
passed the $(NUV-g)_0$ criterion.

\citet{veljanoski13b} recently discovered new GCs in the M31
satellites NGC 147 ($d_\sun =676\pm28$ kpc) and NGC 185 ($d_\sun = 617
\pm 26$ kpc) and provided g- and i-band photometry, which we used for
both the new and the previously known GCs.  Of the 10 GCs now known in
NGC 147, one (SD-GC10) is fainter than our $r_0$ limit and another
(PA-N147-3) had to be excluded from our analysis because the g and i
measurements for it are only lower limits (see Veljanosky et
al. 2013).  Five were identified unambiguously in the WISE catalog
(Hodge I \& III, SD-GC7, PA-N147-1 \& 2)  Two of the remaining GCs (Hodge II \& SD-GC5) appear to be confused with other sources, and
the faint cluster Hodge IV was not recognized by WISE, perhaps because
of the high stellar density in its field (see Veljanosky et al. 2013)
.  All 5 measured clusters (solid squares in Fig. 5) passed the
$(i-W1)_0$ and W2-W3 criteria.  Four of them were identified in the
GALEX catalog, and they all passed the $(NUV-g)_0$ criterion.

In NGC 185 there are now 8 known GCs \citep{veljanoski13b}, and 6
(FJJ-I, III, IV, V, VII, PA-N185) were found in the WISE catalogue.
FJJ-II lies in a dense stellar field, and FJJ-V III appears to be
blended with a nearby source.  All of the 6 clusters with WISE measurements
(open triangles in Fig. 5) passed the $(i-W1)_0$ criterion, but only
5 passed the W2-W3 one.  Two of them were identified in the
GALEX catalog, FJJ-III and FJJ-V, and one, FJJ-III, which lies close to the
center of NGC 185, failed the $(NUV-g)_0$ criterion.

The final dwarf galaxy, NGC 205 ($d_\sun = 824\pm27 $ kpc), is
projected so close to M31 that there is some uncertainty whether a GC
belongs to it or to the M31 halo.  The GCs near the center of NGC 205,
which are among the most likely candidates for membership, are of
little use here because they lie in extremely dense star fields.  We
therefore selected 4 of the more remote clusters, which are also among
the brightest: B009 (Hubble (H)-I), B011 (H-II), B020 (H-III), and
B317 (H-VIII).  The B numbers are from the RBCv5, which provided the
optical photometry.  All 4 passed the criteria based on $(i-W1)_0$,
W2-W3, and $(NUV-g)_0$ (asterisks in Fig. 5).

In summary, of the 42 GCs in the dwarf Local Group galaxies discussed
above, 5 (12\%) are fainter than our $r_0$ cutoff. We could not apply
our survey techniques to another 8 clusters because of the crowding of
images or the lack of g and i photometry. Twenty-nine GCs had adequate
data and 25 (86\%) passed the survey criteria.  If the GCs in these
dwarf galaxies are representative of the IGCs in the LG, then the
majority of these IGCs should pass our selection criteria prior to the
final step of visual inspection.

\section{Results of the Local Group Survey}

As described in section 3, the color selections that are based on the
SDSS, WISE, and GALEX measurements reduced to 183,791 objects the
initial sample of millions of objects from the SDSS Galaxy Catalogue.
This sample is expected to contain a large fraction of the IGCs in
the LG that are brighter than the $r_0$ limit of 19.0, if the M31 GCs
are good proxies for them.  The diagrams in Figure 2 suggest that
$\sim 2 \%$ of the IGCs brighter than the cutoff are missed by the
star-galaxy separation at $d_{\sun} \sim 1100$ kpc and that this
decreases to $\sim 1 \%$ at M31's distance (783 kpc).  Consequently,
the $r_0$ cutoff appears to be the more important of these two
limitations.  It sets a lower limit on the luminosities of the IGCs
that can be detected, which scales, of course, with their distances,
and it is $M_V \leq -6$ at the outer boundary of the LG (see Fig. 2).
The tests with the GCs in M31 (Figs. 3 \& 4) and in the LG dwarf galaxies (Fig. 5) suggest
that a large fraction ($\geq 86 \%$) that are brighter than the $r_0$
cutoff pass the color selection techniques.  It is harder
to quantify the success our visual inspections of the objects
that pass these other criteria, particularly as a function of
distance.  The resolution of the SDSS images, typically $1\farcs4$
seeing with $0\farcs40$ per pixel can make uncertain the distinction
between compact GCs and some galaxies at the distance of M31 and
beyond, even with the improvement of the image quality of the cutouts
provided by SDSS in its DR10 \citep[see][]{ahn14} over previous
releases.  This later improvement would have reduced the larger
misclassifications of candidate GCs that we made in our first survey,
which was based on DR8.  Our present survey retrieved $84\%$ of
the known GCs in M31 and M33 in the search area, which is our best
estimate of the success rate of our methods, including the visual inspection, at $\sim 70\%$ of the radius of the LG.

The visual examination of the SDSS cutout images of the objects that
passed all other criteria yielded a surprisingly small list of 17
potential new GC candidates.  Five of them appeared enough resolved
in the SDSS images that we could classify them as high confidence
candidates.  Because they are near M31, we were able to locate them in
g, r, or i band images taken with the MegaPrime camera of the CFHT and
publicly available through the Canadian Astronomy Data Centre (CADC).
These images are deeper and of higher resolution, typically $0\farcs8$
seeing with $0\farcs19$ per pixel.  With these new images, we were
able to confirm as GCs all 5 of our high confidence objects.  Table 1
provides the positions of these new GCs and the data by which they
were selected.  The SDSS r-band thumbnail images of these GCs are
shown in Figure 6.  These faint GCs may have escaped detection until now
because they are not within the areas covered by Hubble Space
Telescope (HST) images and because they lie at smaller $R_{gc}$ than
the inner boundary ($\sim 25$ kpc) of the PAandAS survey, which
employed MegaPrime imaging.

To take advantage of the better resolution of the DR10 cutout images
and also of the photometric selection criteria that are presented in
this paper, we revisited the lists of high and lower confidence GCs
provided in our first two surveys (Paper I \& II).  Most of the high
confidence objects (tables 1 of Paper I \& II) passed the new
selection criteria, but only three of the lower confidence objects in
table 2 of Paper I passed.  We then visually inspected this reduced sample
through the new cutouts of SDSS DR10.  The ones that we reclassified
as higher confidence candidates could be confirmed as GCs by our
visual inspection of the MegaPrime images from the CADC archive.  Some
of them were later confirmed also by the PAndAS search
\citep{huxor14}.  Table 2 is the updated list of our newly discovered
and confirmed GCs from the SDSS Galaxy Catalogue.  This list is
comprehensive of our 3 surveys and counts 22 GCs, including several
with $R_{gc} > 100$ kpc.  It lists the positions of the clusters, their
reddening corrected values of r and g-i (from the model magnitudes and
extinctions listed in the SDSS), their absolute r magnitudes ($M_r$),
their half-light radii ($R_h$), and their projected distances from M31
($R_{gc}$).  The values of $R_h$ were estimated from the r-band light
profiles provided by the SDSS, and they give rough estimates of the
sizes of the clusters (see Paper I).

The 12 remaining candidates that we could not confirm as GCs are
listed in Table 3, with their positions, photometric data, and angular
sizes of $R_h$.  Their SDSS r-band thumbnail images are shown in
Figure 7.  The distribution on the sky of these 12 candidate GCs is
shown in the Hammer projection of Figure 8.  There is no evidence for
a concentration in a particular region of the sky.  None of them is in
close proximity to the galaxies in or near the LG.  Two of them, C1
and C9, appear in Figure 8 to be close to the dwarf galaxies IC1613
and KKR-25, respectively, but in each case, the angular separation is
$> 3 $ deg and many times the $R_h$ of the galaxy given by
\citet{mcconnachie12}.

\section{Application to the M81 Group of Galaxies}

Our survey area covers many of the galaxies of the M81 group of
galaxies, including M81, M82, and NGC3077, which have interacted with
each other recently and are enclosed in a common envelop of HI gas
\citep{chynoweth08}.  At distance of $\sim 3.6$ Mpc
\citet{karachentsev13}, the M81 group is far from the LG, but
nonetheless our survey techniques can identify the most luminous GCs
as GC candidates.  The two clusters that \citet{jang12} identified in
the images of the HST Advanced Camera for Surveys (ACS), pass all our
photometric criteria (filled triangles in Fig. 9).  Since the M81 GCs
are resolved into stars only on deep images with the HST
\citep[see][]{nantais11}, our visual examination of the SDSS images
could simply reveal if an object was obviously a galaxy or if its
image resembled those of the two clusters in \citet{jang12}.  Over the
area of the sky around the M81 group defined by $143\fdg0 \leq RA \leq
159\fdg0$ and $64\fdg0 \leq DEC \leq 71\fdg 0$, we lowered the $r_0$
limit of our survey to 20.0.  The SDSS footprint covers about 80\% the
region enclosed by the above coordinates and has an odd shape with the
peculiar galaxy M82 near its northern boundary.  Many GCs at the
distance of the M81 group are likely to be indistinguishable from
stars in the SDSS images \citep[see][]{perelmuter95}, and these
clusters will not be listed in the SDSS Galaxy Catalogue.  However, if
the GCs in the M81 group resemble the M31 GCs, $\sim 50\%$ of the most
luminous ones ($M_V \leq -7.8$) will be non-stellar according to the
SDSS criterion (see Fig. 2).  The presence in the Galaxy Catalogue of
the two luminous GCs identified by \citet{jang12} is consistent with
this expectation.  The objects that we identify as GC candidates in
the M81 group are plotted in Figure 9 (open circles) and listed in
table 4, where we have also listed their photometry and projected
distances from M81.  M81-C3, M81-C4, and M81-C5 are closer to the
galaxies M82, NGC3077, and BK6N, respectively, than to M81, and may be
physically related to them.  Two of our candidate GCs, M81-C1 and
M81-C2, are listed as 90262 and 50016 in \citet{perelmuter95}
catalogue of 3774 objects within 25 arcmin of M81.  Neither one
appears to have been investigated since then.  The brighter of the
two, M81-C2, which is also closest to M81, lies within the fields of
2000s and 800s exposures in the F300W pass-band with the HST's
wide-field and planetary camera 2 (WFPC2).  Our inspection of these
images did not reveal any signs that M81-C2 is a galaxy.  The GC 1029
\citep{nantais11}, which is the most luminous one known in M81
\citep[see][]{mayya13}, is very resolved into stars on F814W images
with the HST ACS, but not obviously in a 6300s exposure with the WFPC2
in the F300W filter.  Consequently, the fact that M81-C2 is not
clearly a GC in the F300W WFPC2 images does not rule it out as a
candidate.

\section{Conclusions}

This survey for GCs in the LG has identified 5 new GCs in the halo of
M31 (see Table 1).  Their properties do not appear to be exceptional,
and their discovery suggests that the census of GCs in M31 may be
still incomplete.  The searches of the SDSS Galaxy Catalogue for GCs
described here and in Papers I \& II have yielded a total of 22 new
GCs near M31 (see Table 2).  These clusters span a wide range of
$R_{gc}$ (10-137 kpc), but even the most remote ones are likely to be
members of M31's halo.  It is possible that measurements of their
radial velocities and 3-D distances from M31 could reveal that some
are not bound to it and are IGCs.

Of all the objects that passed our 5 selection criteria for GCs and
our visual inspection, only 12 are so far from M31 that they may be
IGCs in the LG (see Table 3).  While they resemble GCs in the SDSS
imaging (see Fig. 7), there is still the possibility that they are
galaxies, and we consider them to be only GC candidates.  These
objects need to be investigated in more detail to see if they are
truly GCs.

This sample of candidate IGCs in the LG is clearly incomplete because
our survey has covered only about one-third of the sky (see Fig. 8),
and we intend to expand the search in the southern sky.  Moreover,
some IGCs could have escaped our detection because they are hidden by
obscuration near the Galactic Plane, blended with other objects, or
fainter than our magnitude limit, which at the outer reaches of the LG
excludes GCs with $M_V \geq -6$ (see Fig. 2).  Also, as mentioned
previously, about 15\% of the known GCs fail our selection criteria.
Even with these caveats, our survey suggests that the LG does not
appear to have a large population of IGCs, independently of their
possible origin.

\acknowledgments

We gratefully acknowledge the technical support provided by Gabriele
Zinn throughout this project, which greatly facilitated its
completion.  This research has been supported by NSF grant AST-1108948
to Yale University.  This project would not have been possible without
the public release of the data from the Sloan Digital Sky Survey III
and the very useful tools that the SDSS has provided for accessing and
examining the publically released data.  Funding for SDSS-III has been
provided by the Alfred P. Sloan Foundation, the Participating
Institutions, the National Science Foundation, and the U.S. Department
of Energy Office of Science. The SDSS-III web site is
http://www.sdss3.org/.

SDSS-III is managed by the Astrophysical Research Consortium for the
Participating Institutions of the SDSS-III Collaboration including the
University of Arizona, the Brazilian Participation Group, Brookhaven
National Laboratory, University of Cambridge, Carnegie Mellon
University, University of Florida, the French Participation Group, the
German Participation Group, Harvard University, the Instituto de
Astrofisica de Canarias, the Michigan State/Notre Dame/JINA
Participation Group, Johns Hopkins University, Lawrence Berkeley
National Laboratory, Max Planck Institute for Astrophysics, Max Planck
Institute for Extraterrestrial Physics, New Mexico State University,
New York University, Ohio State University, Pennsylvania State
University, University of Portsmouth, Princeton University, the
Spanish Participation Group, University of Tokyo, University of Utah,
Vanderbilt University, University of Virginia, University of
Washington, and Yale University.

This publication makes use of data products from the Wide-field
Infrared Survey Explorer, which is a joint project of the University
of California, Los Angeles, and the Jet Propulsion
Laboratory/California Institute of Technology, funded by the National
Aeronautics and Space Administration.  It also used observations made
by the Galaxy Evolution Explorer satellite, which were obtained
through the GALEXview website.

 This research used the facilities of the Canadian Astronomy Data
 Centre operated by the National Research Council of Canada with the
 support of the Canadian Space Agency.


\begin{thebibliography}{80}
\expandafter\ifx\csname natexlab\endcsname\relax\def\natexlab#1{#1}\fi

\bibitem[{{Abadi} {et~al.}(2003){Abadi}, {Navarro}, {Steinmetz}, \&
  {Eke}}]{abadi03}
{Abadi}, M.~G., {Navarro}, J.~F., {Steinmetz}, M., \& {Eke}, V.~R. 2003, \apj,
  591, 499

\bibitem[{{Abell}(1955)}]{abell55}
{Abell}, G.~O. 1955, \pasp, 67, 258

\bibitem[{{Ahn} {et~al.}(2014){Ahn}, {Alexandroff}, {Allende Prieto}, {Anders},
  {Anderson}, {Anderton}, {Andrews}, {Aubourg}, {Bailey}, {Bastien}, \&
  et~al.}]{ahn14}
{Ahn}, C.~P. {et~al.} 2014, \apjs, 211, 17

\bibitem[{{Alamo-Mart{\'{\i}}nez} {et~al.}(2013){Alamo-Mart{\'{\i}}nez},
  {Blakeslee}, {Jee}, {C{\^o}t{\'e}}, {Ferrarese},
  {Gonz{\'a}lez-L{\'o}pezlira}, {Jord{\'a}n}, {Meurer}, {Peng}, \&
  {West}}]{alamo-mart13}
{Alamo-Mart{\'{\i}}nez}, K.~A. {et~al.} 2013, \apj, 775, 20

\bibitem[{{Ashman} \& {Zepf}(1992)}]{ashman92}
{Ashman}, K.~M., \& {Zepf}, S.~E. 1992, \apj, 384, 50

\bibitem[{{Bate} {et~al.}(2014){Bate}, {Conn}, {McMonigal}, {Lewis}, {Martin},
  {McConnachie}, {Veljanoski}, {Mackey}, {Ferguson}, {Ibata}, {Irwin},
  {Fardal}, {Huxor}, \& {Babul}}]{bate14}
{Bate}, N.~F. {et~al.} 2014, \mnras, 437, 3362

\bibitem[{{Bekki} \& {Yahagi}(2006)}]{bekki06}
{Bekki}, K., \& {Yahagi}, H. 2006, \mnras, 372, 1019

\bibitem[{{Belokurov} {et~al.}(2014){Belokurov}, {Irwin}, {Koposov}, {Evans},
  {Gonzalez-Solares}, {Metcalfe}, \& {Shanks}}]{belokurov14}
{Belokurov}, V., {Irwin}, M.~J., {Koposov}, S.~E., {Evans}, N.~W.,
  {Gonzalez-Solares}, E., {Metcalfe}, N., \& {Shanks}, T. 2014, \mnras, 441,
  2124

\bibitem[{{Bianchi} {et~al.}(2014){Bianchi}, {Conti}, \& {Shiao}}]{bianchi14}
{Bianchi}, L., {Conti}, A., \& {Shiao}, B. 2014, Advances in Space Research,
  53, 900

\bibitem[{{Billett} {et~al.}(2002){Billett}, {Hunter}, \&
  {Elmegreen}}]{billett02}
{Billett}, O.~H., {Hunter}, D.~A., \& {Elmegreen}, B.~G. 2002, \aj, 123, 1454

\bibitem[{{Bland-Hawthorn} \& {Freeman}(2014)}]{bland-hawthorn2014}
{Bland-Hawthorn}, J., \& {Freeman}, K. 2014, The Origin of the Galaxy and Local
  Group, Saas-Fee Advanced Course, Volume 37.~ISBN
  978-3-642-41719-1.~Springer-Verlag Berlin Heidelberg, 2014, p.~1, 37, 1

\bibitem[{{Bullock} \& {Johnston}(2005)}]{bullock05}
{Bullock}, J.~S., \& {Johnston}, K.~V. 2005, \apj, 635, 931

\bibitem[{{Caldwell} {et~al.}(2014){Caldwell}, {Strader}, {Romanowsky},
  {Brodie}, {Moore}, {Diemand}, \& {Martizzi}}]{caldwell14}
{Caldwell}, N., {Strader}, J., {Romanowsky}, A.~J., {Brodie}, J.~P., {Moore},
  B., {Diemand}, J., \& {Martizzi}, D. 2014, \apjl, 787, L11

\bibitem[{{Chynoweth} {et~al.}(2008){Chynoweth}, {Langston}, {Yun}, {Lockman},
  {Rubin}, \& {Scoles}}]{chynoweth08}
{Chynoweth}, K.~M., {Langston}, G.~I., {Yun}, M.~S., {Lockman}, F.~J., {Rubin},
  K.~H.~R., \& {Scoles}, S.~A. 2008, \aj, 135, 1983

\bibitem[{{Cockcroft} {et~al.}(2011){Cockcroft}, {Harris}, {Ferguson}, {Huxor},
  {Ibata}, {Irwin}, {McConnachie}, {Woodley}, {Chapman}, {Lewis}, \&
  {Puzia}}]{cockcroft11}
{Cockcroft}, R. {et~al.} 2011, \apj, 730, 112

\bibitem[{{Cole} {et~al.}(2012){Cole}, {Dehnen}, {Read}, \&
  {Wilkinson}}]{cole12}
{Cole}, D.~R., {Dehnen}, W., {Read}, J.~I., \& {Wilkinson}, M.~I. 2012, \mnras,
  426, 601

\bibitem[{{Conroy} {et~al.}(2011){Conroy}, {Loeb}, \& {Spergel}}]{conroy11}
{Conroy}, C., {Loeb}, A., \& {Spergel}, D.~N. 2011, \apj, 741, 72

\bibitem[{{Conroy} \& {Spergel}(2011)}]{conroyspergel11}
{Conroy}, C., \& {Spergel}, D.~N. 2011, \apj, 726, 36

\bibitem[{{Cutri} {et~al.}(2011){Cutri}, {Wright}, {Conrow}, {Bauer},
  {Benford}, {Brandenburg}, {Dailey}, {Eisenhardt}, {Evans}, {Fajardo-Acosta},
  {Fowler}, {Gelino}, {Grillmair}, {Harbut}, {Hoffman}, {Jarrett},
  {Kirkpatrick}, {Liu}, {Mainzer}, {Marsh}, {Masci}, {McCallon}, {Padgett},
  {Ressler}, {Royer}, {Skrutskie}, {Stanford}, {Wyatt}, {Tholen}, {Tsai},
  {Wachter}, {Wheelock}, {Yan}, {Alles}, {Beck}, {Grav}, {Masiero}, {McCollum},
  {McGehee}, \& {Wittman}}]{cutri11}
{Cutri}, R.~M. {et~al.} 2011, {Explanatory Supplement to the WISE Preliminary
  Data Release Products}, Tech. rep.

\bibitem[{{di Tullio Zinn} \& {Zinn}(2013)}]{ditullio13}
{di Tullio Zinn}, G., \& {Zinn}, R. 2013, \aj, 145, 50

\bibitem[{{di Tullio Zinn} \& {Zinn}(2014)}]{ditullio14}
---. 2014, \aj, 147, 90

\bibitem[{{Elmegreen} {et~al.}(2012){Elmegreen}, {Malhotra}, \&
  {Rhoads}}]{elmegreen12}
{Elmegreen}, B.~G., {Malhotra}, S., \& {Rhoads}, J. 2012, \apj, 757, 9

\bibitem[{{Fardal} {et~al.}(2013){Fardal}, {Weinberg}, {Babul}, {Irwin},
  {Guhathakurta}, {Gilbert}, {Ferguson}, {Ibata}, {Lewis}, {Tanvir}, \&
  {Huxor}}]{fardal13}
{Fardal}, M.~A. {et~al.} 2013, \mnras, 434, 2779

\bibitem[{{Galleti} {et~al.}(2004){Galleti}, {Federici}, {Bellazzini}, {Fusi
  Pecci}, \& {Macrina}}]{galleti04}
{Galleti}, S., {Federici}, L., {Bellazzini}, M., {Fusi Pecci}, F., \&
  {Macrina}, S. 2004, \aap, 416, 917

\bibitem[{{Garrison-Kimmel} {et~al.}(2014){Garrison-Kimmel}, {Boylan-Kolchin},
  {Bullock}, \& {Lee}}]{garrison-kimmel14}
{Garrison-Kimmel}, S., {Boylan-Kolchin}, M., {Bullock}, J.~S., \& {Lee}, K.
  2014, \mnras, 438, 2578

\bibitem[{{Gill} {et~al.}(2005){Gill}, {Knebe}, \& {Gibson}}]{gill05}
{Gill}, S.~P.~D., {Knebe}, A., \& {Gibson}, B.~K. 2005, \mnras, 356, 1327

\bibitem[{{Gordon} \& {Kron}(1983)}]{gordon83}
{Gordon}, K.~C., \& {Kron}, G.~E. 1983, \pasp, 95, 461

\bibitem[{{Gregg} {et~al.}(2009){Gregg}, {Drinkwater}, {Evstigneeva}, {Jurek},
  {Karick}, {Phillipps}, {Bridges}, {Jones}, {Bekki}, \& {Couch}}]{gregg09}
{Gregg}, M.~D. {et~al.} 2009, \aj, 137, 498

\bibitem[{{Harris} \& {Pudritz}(1994)}]{harris94}
{Harris}, W.~E., \& {Pudritz}, R.~E. 1994, \apj, 429, 177

\bibitem[{{Harris} \& {Racine}(1979)}]{harris79}
{Harris}, W.~E., \& {Racine}, R. 1979, \araa, 17, 241

\bibitem[{{Huxor} {et~al.}(2009){Huxor}, {Ferguson}, {Barker}, {Tanvir},
  {Irwin}, {Chapman}, {Ibata}, \& {Lewis}}]{huxor09}
{Huxor}, A., {Ferguson}, A.~M.~N., {Barker}, M.~K., {Tanvir}, N.~R., {Irwin},
  M.~J., {Chapman}, S.~C., {Ibata}, R., \& {Lewis}, G. 2009, \apjl, 698, L77

\bibitem[{{Huxor} {et~al.}(2013){Huxor}, {Ferguson}, {Veljanoski}, {Mackey}, \&
  {Tanvir}}]{huxor13}
{Huxor}, A.~P., {Ferguson}, A.~M.~N., {Veljanoski}, J., {Mackey}, A.~D., \&
  {Tanvir}, N.~R. 2013, \mnras, 429, 1039

\bibitem[{{Huxor} {et~al.}(2014){Huxor}, {Mackey}, {Ferguson}, {Irwin},
  {Martin}, {Tanvir}, {Veljanoski}, {McConnachie}, {Fishlock}, {Ibata}, \&
  {Lewis}}]{huxor14}
{Huxor}, A.~P. {et~al.} 2014, \mnras, 442, 2165

\bibitem[{{Huxor} {et~al.}(2008){Huxor}, {Tanvir}, {Ferguson}, {Irwin},
  {Ibata}, {Bridges}, \& {Lewis}}]{huxor08}
{Huxor}, A.~P., {Tanvir}, N.~R., {Ferguson}, A.~M.~N., {Irwin}, M.~J., {Ibata},
  R., {Bridges}, T., \& {Lewis}, G.~F. 2008, \mnras, 385, 1989

\bibitem[{{Hwang} {et~al.}(2011){Hwang}, {Lee}, {Lee}, {Park}, {Park}, {Kim},
  \& {Park}}]{hwang11}
{Hwang}, N., {Lee}, M.~G., {Lee}, J.~C., {Park}, W.-K., {Park}, H.~S., {Kim},
  S.~C., \& {Park}, J.-H. 2011, \apj, 738, 58

\bibitem[{{Hwang} {et~al.}(2014){Hwang}, {Park}, {Lee}, {Lim}, {Hodge}, {Kim},
  {Miller}, \& {Weisz}}]{hwang14}
{Hwang}, N., {Park}, H.~S., {Lee}, M.~G., {Lim}, S., {Hodge}, P.~W., {Kim},
  S.~C., {Miller}, B., \& {Weisz}, D. 2014, \apj, 783, 49

\bibitem[{{Ibata} {et~al.}(2013){Ibata}, {Nipoti}, {Sollima}, {Bellazzini},
  {Chapman}, \& {Dalessandro}}]{ibata13}
{Ibata}, R., {Nipoti}, C., {Sollima}, A., {Bellazzini}, M., {Chapman}, S.~C.,
  \& {Dalessandro}, E. 2013, \mnras, 428, 3648

\bibitem[{{Jang} {et~al.}(2012){Jang}, {Lim}, {Park}, \& {Lee}}]{jang12}
{Jang}, I.~S., {Lim}, S., {Park}, H.~S., \& {Lee}, M.~G. 2012, \apjl, 751, L19

\bibitem[{{Kang} {et~al.}(2012){Kang}, {Rey}, {Bianchi}, {Lee}, {Kim}, \&
  {Sohn}}]{kang12}
{Kang}, Y., {Rey}, S.-C., {Bianchi}, L., {Lee}, K., {Kim}, Y., \& {Sohn}, S.~T.
  2012, \apjs, 199, 37

\bibitem[{{Karachentsev} {et~al.}(2013){Karachentsev}, {Makarov}, \&
  {Kaisina}}]{karachentsev13}
{Karachentsev}, I.~D., {Makarov}, D.~I., \& {Kaisina}, E.~I. 2013, \aj, 145,
  101

\bibitem[{{Kazantzidis} {et~al.}(2013){Kazantzidis}, {{\L}okas}, \&
  {Mayer}}]{kazantzidis13}
{Kazantzidis}, S., {{\L}okas}, E.~L., \& {Mayer}, L. 2013, \apjl, 764, L29

\bibitem[{{Keller} {et~al.}(2012){Keller}, {Mackey}, \& {Da Costa}}]{keller12}
{Keller}, S.~C., {Mackey}, D., \& {Da Costa}, G.~S. 2012, \apj, 744, 57

\bibitem[{{Koposov} {et~al.}(2007){Koposov}, {de Jong}, {Belokurov}, {Rix},
  {Zucker}, {Evans}, {Gilmore}, {Irwin}, \& {Bell}}]{koposov07}
{Koposov}, S. {et~al.} 2007, \apj, 669, 337

\bibitem[{{Kravtsov} \& {Gnedin}(2005)}]{kravtsov05}
{Kravtsov}, A.~V., \& {Gnedin}, O.~Y. 2005, \apj, 623, 650

\bibitem[{{Laevens} {et~al.}(2014){Laevens}, {Martin}, {Sesar}, {Bernard},
  {Rix}, {Slater}, {Bell}, {Ferguson}, {Schlafly}, {Burgett}, {Chambers},
  {Denneau}, {Draper}, {Kaiser}, {Kudritzki}, {Magnier}, {Metcalfe}, {Morgan},
  {Price}, {Sweeney}, {Tonry}, {Wainscoat}, \& {Waters}}]{laevens14}
{Laevens}, B.~P.~M. {et~al.} 2014, \apjl, 786, L3

\bibitem[{{Law} \& {Majewski}(2010)}]{law10}
{Law}, D.~R., \& {Majewski}, S.~R. 2010, \apj, 718, 1128

\bibitem[{{Lee} {et~al.}(2010){Lee}, {Park}, \& {Hwang}}]{lee10}
{Lee}, M.~G., {Park}, H.~S., \& {Hwang}, H.~S. 2010, Science, 328, 334

\bibitem[{{Mackey} {et~al.}(2010){Mackey}, {Ferguson}, {Irwin}, {Martin},
  {Huxor}, {Tanvir}, {Chapman}, {Ibata}, {Lewis}, \& {McConnachie}}]{mackey10}
{Mackey}, A.~D. {et~al.} 2010, \mnras, 401, 533

\bibitem[{{Mackey} \& {Gilmore}(2003)}]{mackey03}
{Mackey}, A.~D., \& {Gilmore}, G.~F. 2003, \mnras, 340, 175

\bibitem[{{Mackey} \& {Gilmore}(2004)}]{mackey04}
---. 2004, \mnras, 355, 504

\bibitem[{{Madore} \& {Arp}(1979)}]{madore79}
{Madore}, B.~F., \& {Arp}, H.~C. 1979, \apjl, 227, L103

\bibitem[{{Maraston}(1998)}]{maraston98}
{Maraston}, C. 1998, \mnras, 300, 872

\bibitem[{{Maraston}(2005)}]{maraston05}
---. 2005, \mnras, 362, 799

\bibitem[{{Martin} {et~al.}(2006){Martin}, {Ibata}, {Irwin}, {Chapman},
  {Lewis}, {Ferguson}, {Tanvir}, \& {McConnachie}}]{martin06}
{Martin}, N.~F., {Ibata}, R.~A., {Irwin}, M.~J., {Chapman}, S., {Lewis}, G.~F.,
  {Ferguson}, A.~M.~N., {Tanvir}, N., \& {McConnachie}, A.~W. 2006, \mnras,
  371, 1983

\bibitem[{{Martin} {et~al.}(2013{\natexlab{a}}){Martin}, {Schlafly}, {Slater},
  {Bernard}, {Rix}, {Bell}, {Ferguson}, {Finkbeiner}, {Laevens}, {Burgett},
  {Chambers}, {Draper}, {Hodapp}, {Kaiser}, {Kudritzki}, {Magnier}, {Metcalfe},
  {Morgan}, {Price}, {Tonry}, {Wainscoat}, \& {Waters}}]{martin13b}
{Martin}, N.~F. {et~al.} 2013{\natexlab{a}}, \apjl, 779, L10

\bibitem[{{Martin} {et~al.}(2013{\natexlab{b}}){Martin}, {Slater}, {Schlafly},
  {Morganson}, {Rix}, {Bell}, {Laevens}, {Bernard}, {Ferguson}, {Finkbeiner},
  {Burgett}, {Chambers}, {Hodapp}, {Kaiser}, {Kudritzki}, {Magnier}, {Morgan},
  {Price}, {Tonry}, \& {Wainscoat}}]{martin13a}
---. 2013{\natexlab{b}}, \apj, 772, 15

\bibitem[{{Mashchenko} \& {Sills}(2005)}]{mashchenko05}
{Mashchenko}, S., \& {Sills}, A. 2005, \apj, 619, 258

\bibitem[{{Mayer} {et~al.}(2001){Mayer}, {Governato}, {Colpi}, {Moore},
  {Quinn}, {Wadsley}, {Stadel}, \& {Lake}}]{mayer01}
{Mayer}, L., {Governato}, F., {Colpi}, M., {Moore}, B., {Quinn}, T., {Wadsley},
  J., {Stadel}, J., \& {Lake}, G. 2001, \apj, 559, 754

\bibitem[{{Mayya} {et~al.}(2013){Mayya}, {Rosa-Gonz{\'a}lez},
  {Santiago-Cort{\'e}s}, {Rodr{\'{\i}}guez-Merino}, {Vega}, {Torres-Papaqui},
  {Bressan}, \& {Carrasco}}]{mayya13}
{Mayya}, Y.~D., {Rosa-Gonz{\'a}lez}, D., {Santiago-Cort{\'e}s}, M.,
  {Rodr{\'{\i}}guez-Merino}, L.~H., {Vega}, O., {Torres-Papaqui}, J.~P.,
  {Bressan}, A., \& {Carrasco}, L. 2013, \mnras, 436, 2763

\bibitem[{{McConnachie}(2012)}]{mcconnachie12}
{McConnachie}, A.~W. 2012, \aj, 144, 4

\bibitem[{{McLaughlin} \& {van der Marel}(2005)}]{mcLaughlin05}
{McLaughlin}, D.~E., \& {van der Marel}, R.~P. 2005, \apjs, 161, 304

\bibitem[{{Morrissey} {et~al.}(2007){Morrissey}, {Conrow}, {Barlow}, {Small},
  {Seibert}, {Wyder}, {Budav{\'a}ri}, {Arnouts}, {Friedman}, {Forster},
  {Martin}, {Neff}, {Schiminovich}, {Bianchi}, {Donas}, {Heckman}, {Lee},
  {Madore}, {Milliard}, {Rich}, {Szalay}, {Welsh}, \& {Yi}}]{morrissey07}
{Morrissey}, P. {et~al.} 2007, \apjs, 173, 682

\bibitem[{{Nantais} {et~al.}(2011){Nantais}, {Huchra}, {Zezas}, {Gazeas}, \&
  {Strader}}]{nantais11}
{Nantais}, J.~B., {Huchra}, J.~P., {Zezas}, A., {Gazeas}, K., \& {Strader}, J.
  2011, \aj, 142, 183

\bibitem[{{Peebles}(1984)}]{peebles84}
{Peebles}, P.~J.~E. 1984, \apj, 277, 470

\bibitem[{{Peng} {et~al.}(2011){Peng}, {Ferguson}, {Goudfrooij}, {Hammer},
  {Lucey}, {Marzke}, {Puzia}, {Carter}, {Balcells}, {Bridges}, {Chiboucas},
  {del Burgo}, {Graham}, {Guzm{\'a}n}, {Hudson}, {Matkovi{\'c}}, {Merritt},
  {Miller}, {Mouhcine}, {Phillipps}, {Sharples}, {Smith}, {Tully}, \& {Verdoes
  Kleijn}}]{peng11}
{Peng}, E.~W. {et~al.} 2011, \apj, 730, 23

\bibitem[{{Perelmuter} \& {Racine}(1995)}]{perelmuter95}
{Perelmuter}, J.-M., \& {Racine}, R. 1995, \aj, 109, 1055

\bibitem[{{Sales} {et~al.}(2007){Sales}, {Navarro}, {Abadi}, \&
  {Steinmetz}}]{sales07}
{Sales}, L.~V., {Navarro}, J.~F., {Abadi}, M.~G., \& {Steinmetz}, M. 2007,
  \mnras, 379, 1475

\bibitem[{{Samsing}(2015)}]{samsing15}
{Samsing}, J. 2015, \apj, 799, 145

\bibitem[{{Sarajedini} \& {Mancone}(2007)}]{sarajedini07}
{Sarajedini}, A., \& {Mancone}, C.~L. 2007, \aj, 134, 447

\bibitem[{{Schiavon} {et~al.}(2012){Schiavon}, {Caldwell}, {Morrison},
  {Harding}, {Courteau}, {MacArthur}, \& {Graves}}]{schiavon12}
{Schiavon}, R.~P., {Caldwell}, N., {Morrison}, H., {Harding}, P., {Courteau},
  S., {MacArthur}, L.~A., \& {Graves}, G.~J. 2012, \aj, 143, 14

\bibitem[{{Schweizer}(1987)}]{schweizer87}
{Schweizer}, F. 1987, in Nearly Normal Galaxies. From the Planck Time to the
  Present, ed. S.~M. {Faber}, 18--25

\bibitem[{{Stoughton} {et~al.}(2002){Stoughton}, {Lupton}, {Bernardi},
  {Blanton}, {Burles}, {Castander}, {Connolly}, {Eisenstein}, {Frieman},
  {Hennessy}, {Hindsley}, {Ivezi{\'c}}, {Kent}, {Kunszt}, {Lee}, {Meiksin},
  {Munn}, {Newberg}, {Nichol}, {Nicinski}, {Pier}, {Richards}, {Richmond},
  {Schlegel}, {Smith}, {Strauss}, {SubbaRao}, {Szalay}, {Thakar}, {Tucker},
  {Vanden Berk}, {Yanny}, {Adelman}, {Anderson}, {Anderson}, {Annis},
  {Bahcall}, {Bakken}, {Bartelmann}, {Bastian}, {Bauer}, {Berman},
  {B{\"o}hringer}, {Boroski}, {Bracker}, {Briegel}, {Briggs}, {Brinkmann},
  {Brunner}, {Carey}, {Carr}, {Chen}, {Christian}, {Colestock}, {Crocker},
  {Csabai}, {Czarapata}, {Dalcanton}, {Davidsen}, {Davis}, {Dehnen},
  {Dodelson}, {Doi}, {Dombeck}, {Donahue}, {Ellman}, {Elms}, {Evans}, {Eyer},
  {Fan}, {Federwitz}, {Friedman}, {Fukugita}, {Gal}, {Gillespie}, {Glazebrook},
  {Gray}, {Grebel}, {Greenawalt}, {Greene}, {Gunn}, {de Haas}, {Haiman},
  {Haldeman}, {Hall}, {Hamabe}, {Hansen}, {Harris}, {Harris}, {Harvanek},
  {Hawley}, {Hayes}, {Heckman}, {Helmi}, {Henden}, {Hogan}, {Hogg}, {Holmgren},
  {Holtzman}, {Huang}, {Hull}, {Ichikawa}, {Ichikawa}, {Johnston}, {Kauffmann},
  {Kim}, {Kimball}, {Kinney}, {Klaene}, {Kleinman}, {Klypin}, {Knapp},
  {Korienek}, {Krolik}, {Kron}, {Krzesi{\'n}ski}, {Lamb}, {Leger},
  {Limmongkol}, {Lindenmeyer}, {Long}, {Loomis}, {Loveday}, {MacKinnon},
  {Mannery}, {Mantsch}, {Margon}, {McGehee}, {McKay}, {McLean}, {Menou},
  {Merelli}, {Mo}, {Monet}, {Nakamura}, {Narayanan}, {Nash}, {Neilsen},
  {Newman}, {Nitta}, {Odenkirchen}, {Okada}, {Okamura}, {Ostriker}, {Owen},
  {Pauls}, {Peoples}, {Peterson}, {Petravick}, {Pope}, {Pordes}, {Postman},
  {Prosapio}, {Quinn}, {Rechenmacher}, {Rivetta}, {Rix}, {Rockosi}, {Rosner},
  {Ruthmansdorfer}, {Sandford}, {Schneider}, {Scranton}, {Sekiguchi}, {Sergey},
  {Sheth}, {Shimasaku}, {Smee}, {Snedden}, {Stebbins}, {Stubbs}, {Szapudi},
  {Szkody}, {Szokoly}, {Tabachnik}, {Tsvetanov}, {Uomoto}, {Vogeley}, {Voges},
  {Waddell}, {Walterbos}, {Wang}, {Watanabe}, {Weinberg}, {White}, {White},
  {Wilhite}, {Wolfe}, {Yasuda}, {York}, {Zehavi}, \& {Zheng}}]{Stoughton02}
{Stoughton}, C. {et~al.} 2002, \aj, 123, 485

\bibitem[{{Teyssier} {et~al.}(2012){Teyssier}, {Johnston}, \&
  {Kuhlen}}]{teyssier12}
{Teyssier}, M., {Johnston}, K.~V., \& {Kuhlen}, M. 2012, \mnras, 426, 1808

\bibitem[{{Veljanoski} {et~al.}(2013){Veljanoski}, {Ferguson}, {Huxor},
  {Mackey}, {Fishlock}, {Irwin}, {Tanvir}, {Chapman}, {Ibata}, {Lewis}, \&
  {McConnachie}}]{veljanoski13b}
{Veljanoski}, J. {et~al.} 2013, \mnras, 435, 3654

\bibitem[{{Veljanoski} {et~al.}(2014){Veljanoski}, {Mackey}, {Ferguson},
  {Huxor}, {C{\^o}t{\'e}}, {Irwin}, {Tanvir}, {Pe{\~n}arrubia}, {Bernard},
  {Fardal}, {Martin}, {McConnachie}, {Lewis}, {Chapman}, {Ibata}, \&
  {Babul}}]{veljanoski14}
---. 2014, \mnras, 442, 2929

\bibitem[{{West} \& {Gregg}(2014)}]{west14}
{West}, M., \& {Gregg}, M. 2014, in American Astronomical Society Meeting
  Abstracts, Vol. 223, American Astronomical Society Meeting Abstracts,
  \#106.03

\bibitem[{{West} {et~al.}(2011){West}, {Jord{\'a}n}, {Blakeslee},
  {C{\^o}t{\'e}}, {Gregg}, {Takamiya}, \& {Marzke}}]{west11}
{West}, M.~J., {Jord{\'a}n}, A., {Blakeslee}, J.~P., {C{\^o}t{\'e}}, P.,
  {Gregg}, M.~D., {Takamiya}, M., \& {Marzke}, R.~O. 2011, \aap, 528, A115

\bibitem[{{Whitmore} \& {Schweizer}(1995)}]{whitmore95}
{Whitmore}, B.~C., \& {Schweizer}, F. 1995, \aj, 109, 960

\bibitem[{{Zinn}(1993)}]{zinn1993}
{Zinn}, R. 1993, in Astronomical Society of the Pacific Conference Series,
  Vol.~48, The Globular Cluster-Galaxy Connection, ed. G.~H. {Smith} \& J.~P.
  {Brodie}, 38

\bibitem[{{Zolotov} {et~al.}(2009){Zolotov}, {Willman}, {Brooks}, {Governato},
  {Brook}, {Hogg}, {Quinn}, \& {Stinson}}]{zolotov09}
{Zolotov}, A., {Willman}, B., {Brooks}, A.~M., {Governato}, F., {Brook}, C.~B.,
  {Hogg}, D.~W., {Quinn}, T., \& {Stinson}, G. 2009, \apj, 702, 1058

\end{thebibliography}

\clearpage

\begin{figure}
\epsscale{0.5}
\plotone{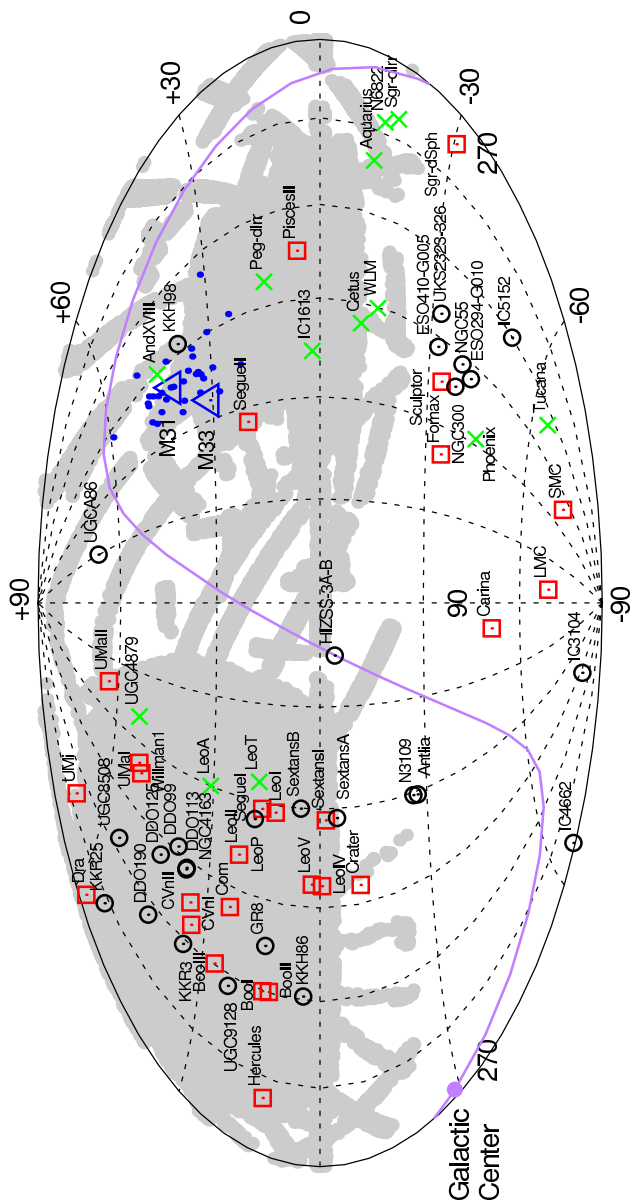}
\caption{A Hammer projection of the sky with equatorial coordinates
  showing the area of our survey (gray shading), the Galactic Center
  (solid circle) and Plane (solid curve), M31 and M33 (open
  triangles), MW satellites (squares), M31 satellites (dots), the other LG
  galaxies (X's), and galaxies near the LG (circles).}
\end{figure}

\begin{figure}
\epsscale{0.5}
\plotone{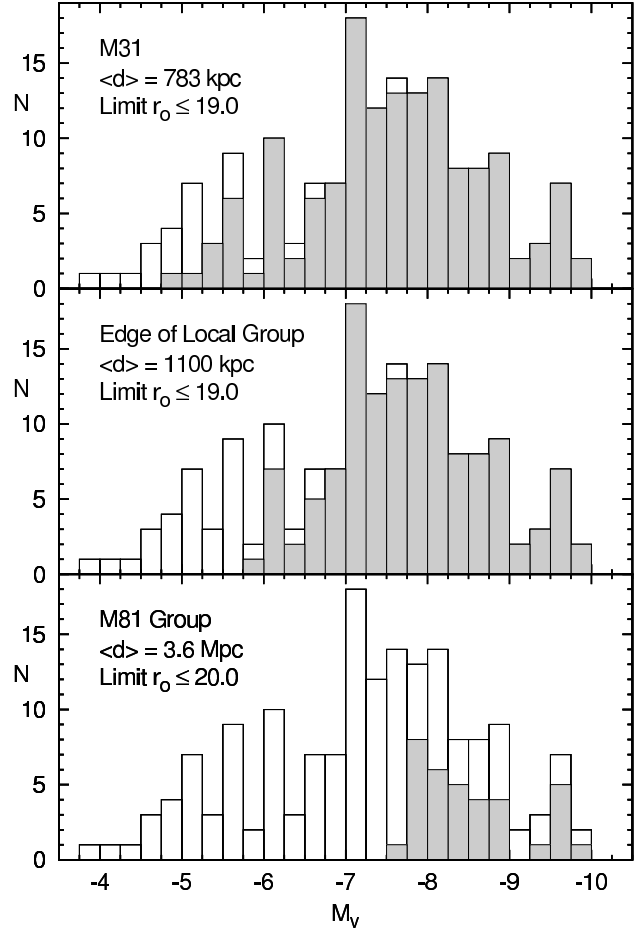}
\caption{The effects of the star-galaxy separation and the magnitude
  limit on the surveys is illustrated using a sample on 168 GCs in
  M31.  The open histogram in each diagram is the luminosity
  distribution of the whole sample.  The solid histograms are the GCs
  that are non-stellar according to the criterion used by the SDSS
  (see text) and are brighter than the listed $ r_{0}$ limit, at the
  assumed mean distance (\textless d\textgreater).  Top: M31's
  distance, middle: the outer boundary of the LG, and bottom: the
  distance of the M81 group.}
\end{figure}

\begin{figure}
\epsscale{0.5}
\plotone{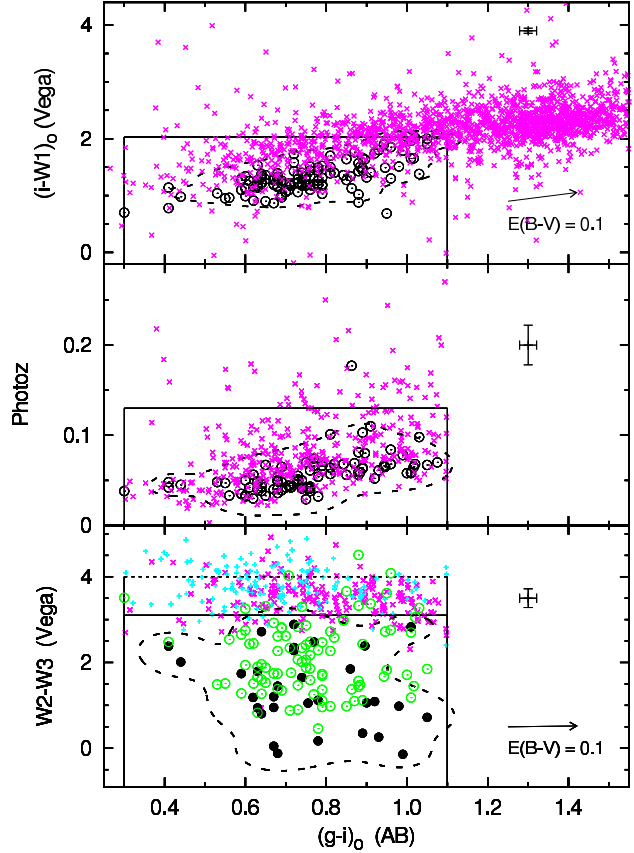}
\caption{The selection criteria that were used in the automated part
  of our survey are illustrated using 122 GCs in the halo of M31
  (circles) and the galaxies (X's) in a $4\: \mathrm{deg}^2$ test region
  of our LG survey.  The dashed contours in each diagram enclose 90\%
  of the 2-D density of the GCs (see text). The whole sample of GC's
  is plotted in each diagram.  The whole sample of galaxies is
  plotted in the top diagram, but only the ones within the rectangle
  in the top diagram are plotted in the middle diagram.  Likewise,
  only the galaxies within the rectangle in the middle diagram are
  plotted in the bottom diagram. In the bottom diagram, the solid
  circles and the X's represent GCs and galaxies, respectively, that
  have measured W2-W3, while the open circles and crosses are the
  upper limits on W2-W3 for other GCs and galaxies, respectively.  The
  lines at W2-W3 = 3.1 (solid) and 4.0 (dashed) are the cuts for
  measured and upper limit values, respectively.  Note that Photoz is
  used as a color index and not a measure of redshift for the GCs.}
\end{figure}

\begin{figure}
\epsscale{0.6}
\plotone{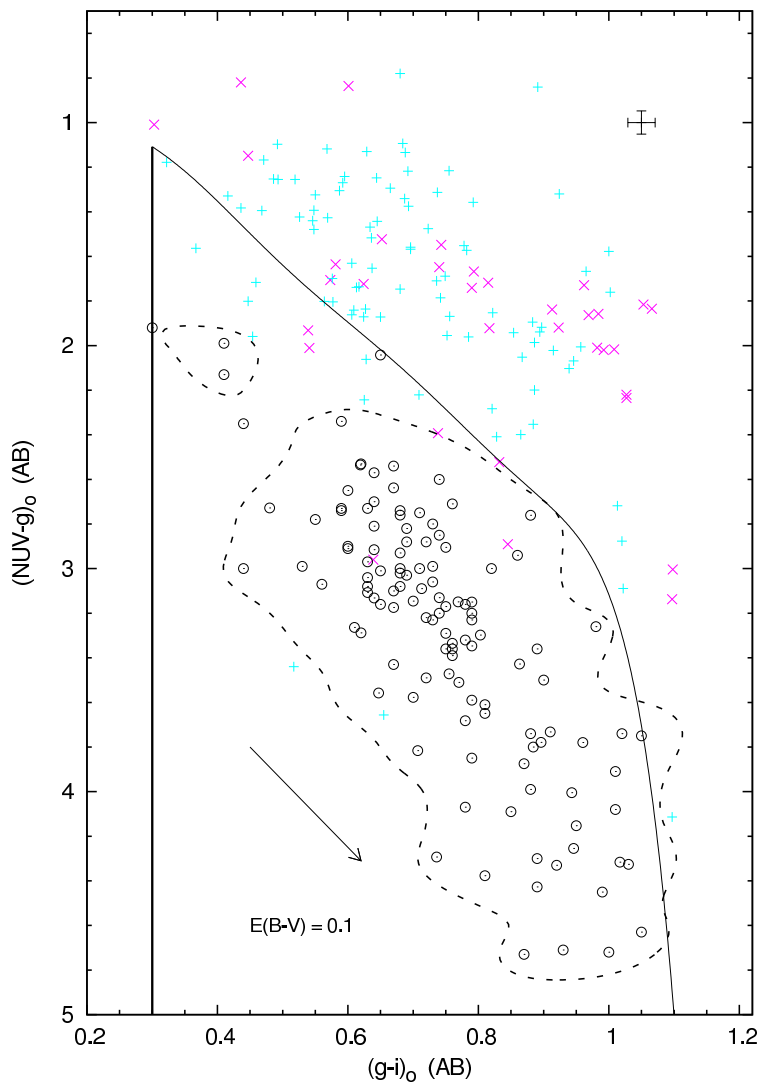}
\caption{The sample of 122 M31 GCs (open circles) are used to
  illustrate the criterion that is based on the color formed by the
  GALEX NUV and the SDSS g magnitudes.  The dashed contour encloses
  90\% of the density of GCs.  The X's and crosses are the galaxies
  that passed the cuts for measured and upper limit values of W2-W3 in
  Figure 3.  The area enclosed by the curve and the vertical lines
  define the criterion used to select GC candidates.}
\end{figure}

\begin{figure}
\epsscale{0.60}
\plotone{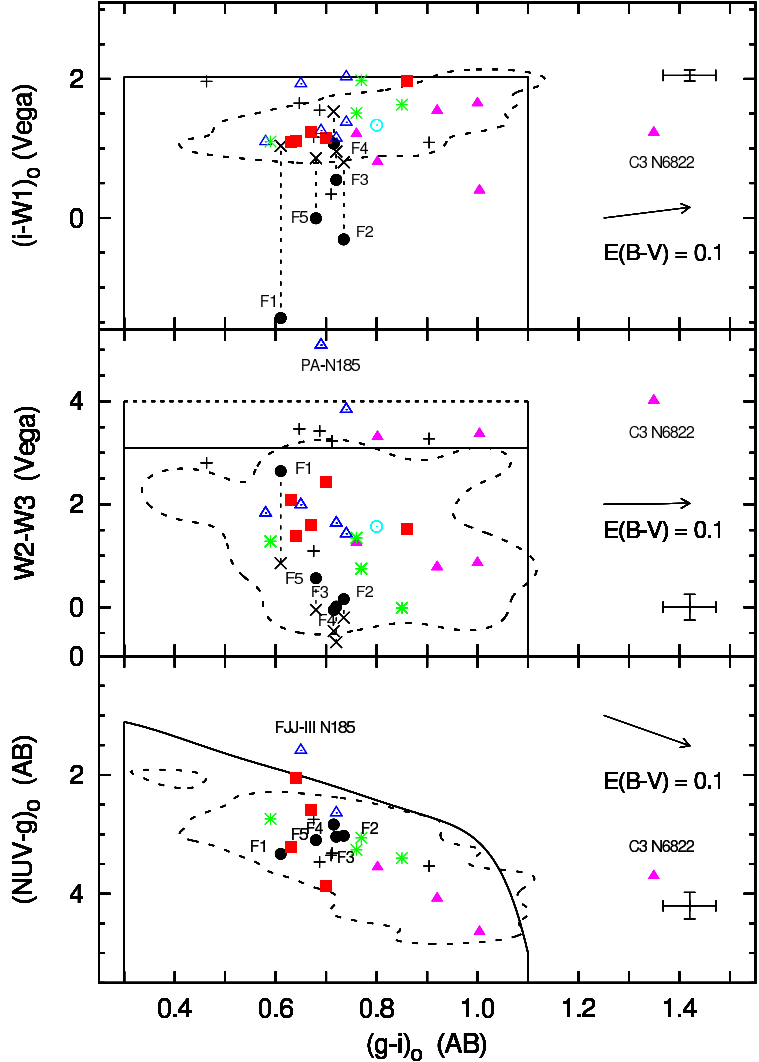}
\caption{The GCs in LG dwarf galaxies are plotted in the diagrams used
  to select GC candidates.  The GCs in the Fornax dSph galaxy, which
  are labeled, are plotted with the WISE profile fitted magnitudes
  (solid circles) and the magnitude in the largest WISE aperture
  (X's), with a dashed line connecting the two.  WISE profile fitted
  magnitudes are used for GCs in NGC 6822 (solid triangles), WLM (open
  circle), M33 (crosses), NGC 147 (solid squares), NGC 185 (open
  triangles), and NGC 205 (asterisks).  The dashed contours are the
  same as the ones plotted in Figures 3 and 4, and they enclose 90\%
  of the density of M31 GCs.}
\end{figure}

\begin{figure}
\epsscale{0.80}
\plotone{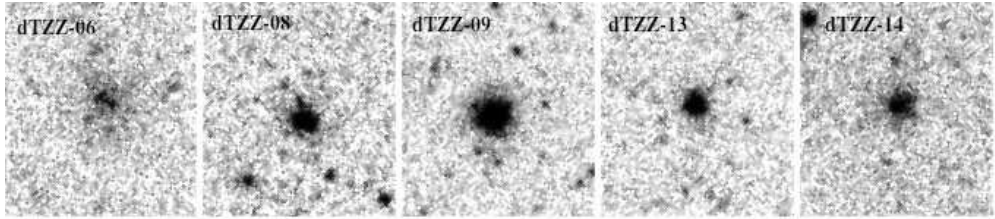}
\caption{The r-band SDSS images of the newly discovered GCs in M31
  (see Table 1).  North is at the top, and East is to the left.  Each
  image is approximately 25'' by 30''.}
\end{figure}

\begin{figure}
\epsscale{0.80}
\plotone{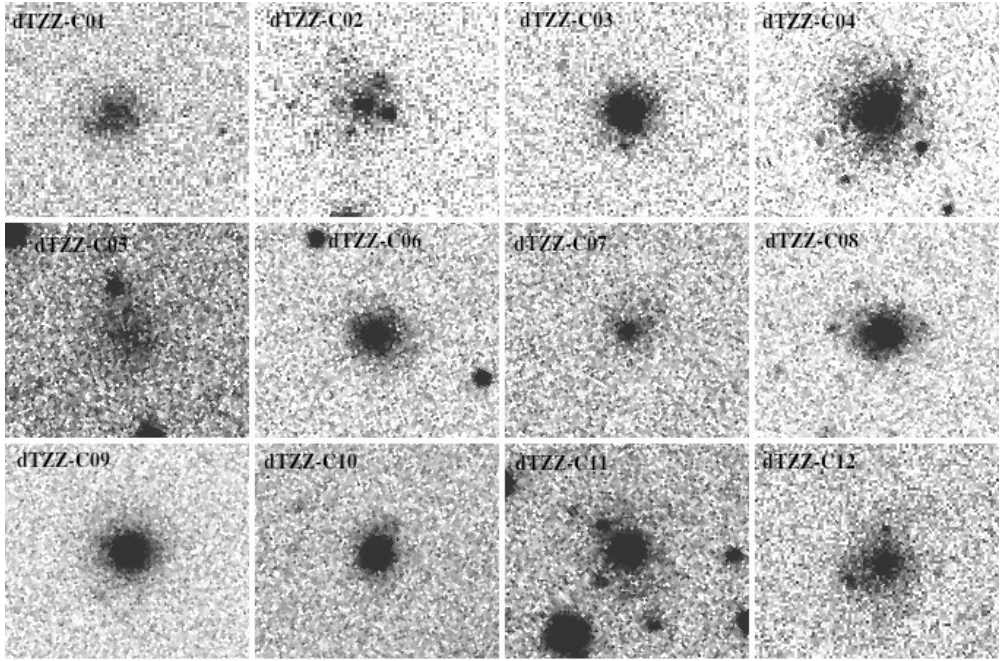}
\caption{The r-band SDSS images of the candidate IGCs in the Local
  Group (see Table 3).  North is at the top, and East is to the left.
  Each image is approximately 35'' by 30''.}
\end{figure}

\begin{figure}
\epsscale{0.60}
\plotone{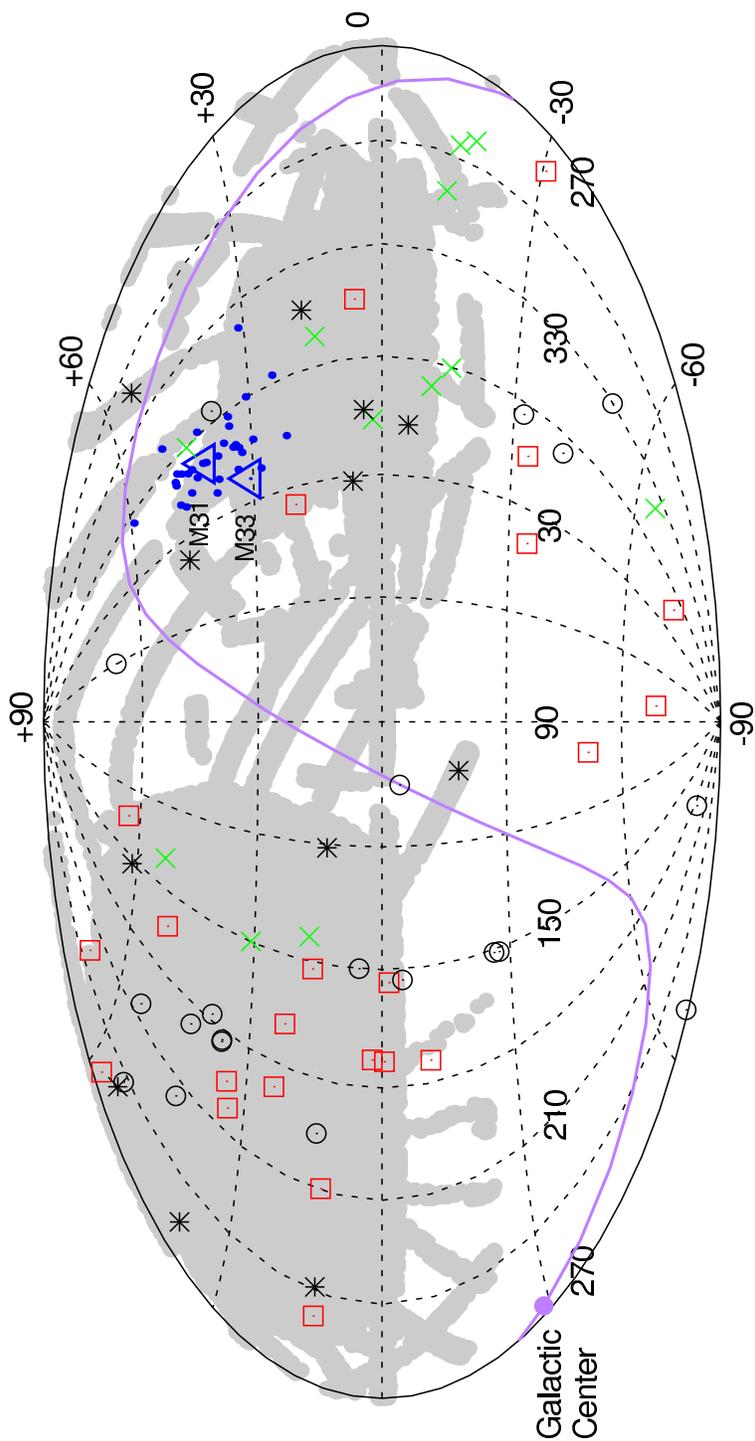}
\caption{A Hammer projection of the sky with equatorial coordinates.
  The candidate IGCs in the LG are plotted as asterisks.  The rest of
  the symbols are the same as Figure 1.} 
\end{figure}

\begin{figure}
\epsscale{0.80}
\plotone{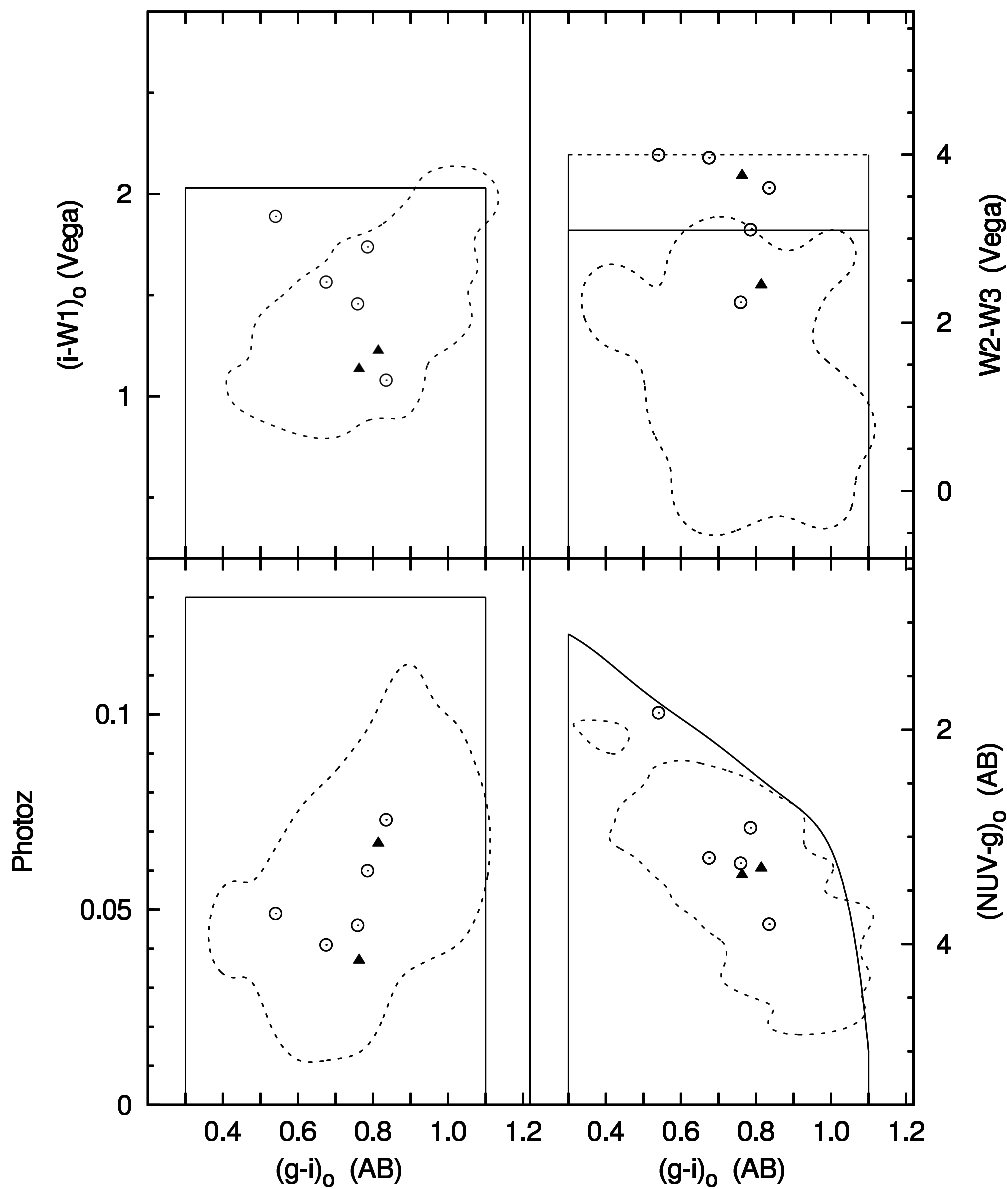}
\caption{The candidate GCs in the M81 Group (open circles) and the two
  GCs identified by Jang et al (2012) (solid triangles) are plotted in
  the 4 diagrams that are used to distinguish GC candidates from
  galaxies (see Figures 3 \& 4).  The dashed contours are they same
  ones that are plotted in Figures 3, 4, \& 5, and they enclose 90\%
  of the density of M31 GCs.}
\end{figure}



\clearpage

\begin{deluxetable}{cccccccccc}

\rotate
\tabletypesize{\small}

\tablewidth{600pt}

\tablecaption{New M31 Globular Clusters}

\tablenum{1}

\tablehead{\colhead{Name} & \colhead{R.A.} & \colhead{Decl.} & \colhead{$r_0$} &
 \colhead{$(NUV-g)_0$} & \colhead{$(g-i)_0$} & \colhead{$(i-W1)_0$} & \colhead{$
W2-W3$\tablenotemark{a}} & \colhead{Photoz\tablenotemark{b}} & \colhead{} \\ 
\colhead{} & \colhead{(deg J2000)} & \colhead{(deg J2000)} & \colhead{(mag)} & \colhead{(mag)} & \colhead{(mag)} & \colhead{(mag)} & \colhead{(mag)} & \colhead{} & \colhead{} } 

\startdata
dTZZ-06 & 9.48287 & 41.30932 & 18.82$\pm$0.03 & 2.62$\pm$0.08 & 0.72$\pm$0.06 & 1.11$\pm$0.15 & 3.63 & 0.043$\pm$0.032 \\
dTZZ-08 & 9.63313 & 39.86855 & 18.22$\pm$0.01 & 4.69$\pm$0.27 & 1.01$\pm$0.02 & 1.59$\pm$0.06 & 3.28 & 0.088$\pm$0.051 \\
dTZZ-09 & 9.75310 & 41.23024 & 17.89$\pm$0.01 & 3.89$\pm$0.07 & 0.81$\pm$0.03 & 1.04$\pm$0.08 & 3.54 & 0.119$\pm$0.046 \\
dTZZ-13 & 11.43322 & 42.65674 & 18.90$\pm$0.02 & 3.84$\pm$0.16 & 1.04$\pm$0.03 & 2.02$\pm$0.07 & 3.79 & 0.050$\pm$0.019 \\
dTZZ-14 & 11.59881 & 42.60960 & 18.92$\pm$0.02 & \nodata & 1.02$\pm$0.04 & 1.05$\pm$0.14 & 3.62 & 0.078$\pm$0.034 &  \\
\enddata
\tablecomments{$(i-W1)_0$ and $W2-W3$ are on the Vega system.  All other magnitudes are on the AB system.}
\tablenotetext{a}{Upper limits}
\tablenotetext{b}{Photoz is used as a color index, not as a measure of redshift.
}

\end{deluxetable}

\clearpage

\begin{deluxetable}{cccccccccc}

\tabletypesize{\scriptsize}

\tablecaption{Updated List of Newly Discovered M31 GCs in the SDSS Catalogue}

\tablenum{2}

\tablehead{\colhead{Cluster} & \colhead{Previous\tablenotemark{a}} & \colhead{R.
A.} & \colhead{Decl.} & \colhead{$r_0$} & \colhead{$(g-i)_0$} & \colhead{$M_r$} 
& \colhead{$R_h$} & \colhead{$R_{gc}$} & \colhead{Confirmation} \\ 
\colhead{Name} & \colhead{Name} & \colhead{(deg J2000)} & \colhead{(deg J2000)} 
& \colhead{(mag)} & \colhead{(mag)} & \colhead{(mag)} & \colhead{(pc)} & \colhead{(kpc)} & \colhead{} } 

\startdata
dTZZ-01 & A & 5.14119 & 36.65953 & 17.44 & 0.82 & -7.0 & 8.5 & 86 & b.c \\
dTZZ-02 & B & 6.71750 & 38.74947 & 15.99 & 0.69 & -8.5 & 4.2 & 54 & b,c \\
dTZZ-03 & C & 7.86467 & 39.53942 & 17.33 & 0.68 & -7.1 & 4.1 & 38 & b,c \\
dTZZ-04 & SDSS1 & 9.00774 & 40.49723 & 17.36 & 1.08 & -7.1 & 10.5 & 20 & b,c \\
dTZZ-05 & D & 9.03580 & 39.29165 & 17.24 & 0.73 & -7.2 & 4.1 & 32 & b \\
dTZZ-06 & \nodata & 9.48287 & 41.30932 & 18.82 & 0.72 & -5.6 & 9.7 & 12 & b \\
dTZZ-07 & E & 9.61483 & 40.65835 & 18.26 & 1.01 & -6.2 & 5.6 & 14 & b \\
dTZZ-08 & \nodata & 9.63313 & 39.86855 & 18.22 & 1.01 & -6.2 & 5.1 & 22 & b \\
dTZZ-09 & \nodata & 9.75310 & 41.23024 & 17.89 & 0.81 & -6.6 & 6.3 & 10 & b \\
dTZZ-10 & SDSS3 & 9.80443 & 41.70220 & 18.25 & 1.08 & -6.2 & 5.9 & 11 & b,c \\
dTZZ-11 & SDSS4 & 10.32496 & 42.77124 & 17.71 & 1.14 & -6.8 & 7.2 & 21 & b,c \\
dTZZ-12 & SDSS6 & 10.61489 & 39.92444 & 18.51 & 0.93 & -6.0 & 7.1 & 18 & b,c \\
dTZZ-13 & \nodata & 11.43322 & 42.65674 & 18.90 & 1.04 & -5.6 & 4.7 & 20 & b \\
dTZZ-14 & \nodata & 11.59881 & 42.60960 & 18.92 & 1.02 & -5.5 & 6.3 & 21 & b \\
dTZZ-15 & SDSS8 & 12.65142 & 42.53047 & 18.57 & 1.10 & -5.9 & 9.1 & 26 & b,c \\
dTZZ-16 & SDSS9 & 13.41490 & 42.58747 & 17.20 & 0.66 & -7.3 & 7.2 & 33 & b,c \\
dTZZ-17 & SDSS11 & 14.73495 & 42.46061 & 15.61 & 0.70 & -8.9 & 4.0 & 44 & b,c \\
dTZZ-18 & SDSS12 & 18.19590 & 42.42356 & 16.88 & 0.74 & -7.6 & 5.0 & 78 & b,c \\
dTZZ-19 & SDSS15 & 20.76470 & 41.91971 & 16.77 & 0.70 & -7.7 & 4.0 & 103 & b,c \\
dTZZ-20 & C62 & 21.94838 & 40.67996 & 18.71 & 0.76 & -5.8 & 11.3 & 116 & b,c \\
dTZZ-21 & G & 22.20478 & 47.07277 & 16.98 & 0.81 & -7.5 & 6.1 & 137 & b,d \\
dTZZ-22 & SDSS16 & 22.25898 & 40.78570 & 18.25 & 0.83 & -6.2 & 6.6 & 119 & b,c \\
\enddata
\tablenotetext{a}{Name used in Paper I or II.}
\tablecomments{Confirmation key: b = our visual inspection of MegaPrime images, 
c = Huxor et al. (2014), d = Huxor, A. (private communication).}

\end{deluxetable}

\begin{deluxetable}{cccccccccc}

\rotate

\tabletypesize{\scriptsize}

\tablewidth{550pt}

\tablecaption{Candidate Intergalactic Globular Clusters in the Local Group}

\tablenum{3}

\tablehead{\colhead{Name} & \colhead{R.A.} & \colhead{Decl.} & \colhead{$r_0$} & \colhead{$(NUV-g)_0$} & \colhead{$(g-i)_0$} & \colhead{$(i-W1)_0$} & \colhead{$(W2-W3$} & \colhead{Photoz} & \colhead{$R_{h}$}\\ 
\colhead{} & \colhead{(deg J2000)} & \colhead{(deg J2000)} & \colhead{(mag)} & \colhead{(mag)} & \colhead{(mag)} & \colhead{(mag)} & \colhead{(mag)} & \colhead{} & \colhead{(arcsec)}} 

\startdata
dTZZ-C01 & 13.61358 & 4.18373 & 18.72$\pm$0.08 & 1.68$\pm$0.06 & 0.31$\pm$0.12 & 1.46$\pm$0.17 & 2.78\tablenotemark{a} & 0.041$\pm$0.025 & 2.7 \\
dTZZ-C02 & 17.34474 & -5.91597 & 18.08$\pm$0.04 & \nodata & 0.73$\pm$0.04 & 1.67$\pm$0.06 & 2.82\tablenotemark{a} & 0.072$\pm$0.075 & 3.4 \\
dTZZ-C03 & 31.37672 & 6.77809 & 17.84$\pm$0.01 & 1.98$\pm$0.14$\pm$ & 0.46$\pm$0.02 & 1.31$\pm$0.06 & 3.30\tablenotemark{a} & 0.034$\pm$0.010 & 2.1 \\
dTZZ-C04 & 37.649609 & 46.31996 & 17.25$\pm$0.01 & 2.29$\pm$0.20 & 0.85$\pm$0.02 & 1.22$\pm$0.05 & 3.02$\pm$0.33 & 0.071$\pm$0.041 & 3.2 \\
dTZZ-C05 & 102.15198 & -18.38880 & 17.71$\pm$0.04 & \nodata & 0.69$\pm$0.07 & 1.19$\pm$0.10 & 3.84\tablenotemark{a} & 0.100$\pm$0.057 & 2.7: \\
dTZZ-C06 & 120.87215 & 13.07730 & 18.20$\pm$0.01 & \nodata & 0.85$\pm$0.03 & 1.70$\pm$0.07 & 3.95\tablenotemark{a} & 0.067$\pm$0.029 & 2.2 \\
dTZZ-C07 & 152.40421 & 61.26622 & 18.53$\pm$0.03 & 2.01$\pm$0.07 & 0.60$\pm$0.05 & 0.19$\pm$0.20 & 3.57\tablenotemark{a} & 0.047$\pm$0.037 & 2.1 \\
dTZZ-C08 & 238.65523 & 12.92038 & 18.00$\pm$0.01 & 1.47$\pm$0.10 & 0.37$\pm$0.03 & 0.91$\pm$0.11 & 3.94\tablenotemark{a} & 0.035$\pm$0.019 & 2.1 \\
dTZZ-C09 & 250.07475 & 54.96822 & 17.66$\pm$0.01 & 3.78$\pm$0.45 & 0.93$\pm$0.01 & 1.41$\pm$0.03 & 2.30\tablenotemark{a} & 0.074$\pm$0.036 & 2.4 \\
dTZZ-C10 & 255.93616 & 38.79727 & 18.02$\pm$0.01 & 2.89$\pm$0.08 & 0.93$\pm$0.02 & 1.90$\pm$0.04 & 2.58$\pm$0.39 & 0.068$\pm$0.028 & 1.2 \\
dTZZ-C11 & 313.82498 & 54.71283 & 15.51$\pm$0.01 & \nodata & 0.85$\pm$0.02 & 1.56$\pm$0.03 & 1.14\tablenotemark{a} & 0.079$\pm$0.051 & 1.3 \\
dTZZ-C12 & 343.69748 & 17.43929 & 18.08$\pm$0.02 & \nodata & 0.63$\pm$0.03 & 1.36$\pm$0.08 & 3.90\tablenotemark{a} & 0.067$\pm$0.016 & 2.8 \\
\enddata

\tablenotetext{a}{Upper limit}

\tablecomments{Photoz is used as a color index and not a measure of redshift.}

\end{deluxetable}

\begin{deluxetable}{cccccccccc}

\rotate

\tabletypesize{\scriptsize}

\tablewidth{550pt}

\tablecaption{Candidates and Globular Clusters in the M81 Group}

\tablenum{4}

\tablehead{\colhead{Name} & \colhead{R.A.} & \colhead{Decl.} & \colhead{$r_0$} & \colhead{$(NUV-g)_0$} & \colhead{$(g-i)_0$} & \colhead{$(i-W1)_0$} & \colhead{$W2-W3$\tablenotemark{a}} & \colhead{Photoz} & \colhead{$R_{gc}$\tablenotemark{b}} \\ 
\colhead{} & \colhead{(deg J2000)} & \colhead{(deg J2000)} & \colhead{(mag)} & \colhead{(mag)} & \colhead{(mag)} & \colhead{(mag)} & \colhead{(mag)} & \colhead{} & \colhead{(kpc)} } 

\startdata
M81-C1 & 148.10805 & 68.80711 & 19.22$\pm$0.02 & 1.84$\pm$0.03  & 0.54$\pm$0.03 & 1.89$\pm$0.10 & 3.99 & 0.049$\pm$0.014 & 24 \\
M81-C2 & 148.57085 & 68.92275 & 17.37$\pm$0.02 & 3.24$\pm$0.03  & 0.76$\pm$0.01 & 1.46$\pm$0.04 & 2.24 & 0.046$\pm$0.024 & 12 \\
M81-C3 & 149.52321 & 69.57946 & 18.46$\pm$0.01 & 3.20$\pm$0.04  & 0.68$\pm$0.02 & 1.57$\pm$0.08 & 3.96 & 0.041$\pm$0.026 & 36 \\
M81-C4 & 151.32164 & 68.77572 & 18.51$\pm$0.01 & 3.82$\pm$0.32  & 0.84$\pm$0.02 & 1.08$\pm$0.11 & 3.61 & 0.073$\pm$0.031 & 58 \\
M81-C5 & 158.17458 & 65.70965 & 17.99$\pm$0.01 & 2.92$\pm$0.10  & 0.79$\pm$0.01 & 1.74$\pm$0.04 & 3.11 & 0.060$\pm$0.020 & 309 \\
GC-1\tablenotemark{c} & 148.35931 & 69.52164 & 18.26$\pm$0.01 & 3.36$\pm$0.05 & 0.76$\pm$0.02 & 1.14$\pm$0.09 & 3.75 & 0.037$\pm$0.021 & 31 \\
GC-2\tablenotemark{c} & 148.33411 & 69.65462 & 17.31$\pm$0.01 & 3.29$\pm$0.03 & 0.81$\pm$0.01 & 1.23$\pm$0.05 & 2.45 & 0.067$\pm$0.030 & 39 \\
\enddata

\tablenotetext{a}{Upper limit}
\tablenotetext{b}{Projected distance from M81.}
\tablenotetext{c}{Known globular cluster (Jang et al. 2012)}

\tablecomments{Photoz is used as a color index and not a measure of redshift.}

\end{deluxetable}

\end{document}